\documentclass[traditabstract]{aa} % for the abstract without structuration 

\usepackage{graphicx}
\usepackage{txfonts}
\usepackage{natbib}

\def\kms{$\rm \,km\,s^{-1}$}

\def\HH{H$_2$}
\def\msol{M$_{\odot}$}

\def\Ha{H${\alpha}$}

\def\Bd{Br${\delta}$}
\def\Bg{Br${\gamma}$}
\def\fwhm{{\tiny FWHM}}

\def\micro{$\rm \,\mu m$}

\def\flux{$\rm \,erg\,s^{-1}\,cm^{-2}$}

\def\HII{\ion{H}{ii}}
\def\ArIII{\ion{Ar}{III}}
\def\SIV{\ion{S}{IV}}
\def\NeII{\ion{Ne}{II}}

\def\2nd{2$^{\rm nd}$}
\def\3rd{3$^{\rm rd}$}
\def\4th{4$^{\rm th}$}

\begin{document}
   \title{High angular resolution near-infrared integral field
     observations of young star cluster complexes in NGC1365}

   \author{E. Galliano\inst{1}
          \and 
          M.Kissler-Patig\inst{2}
          \and
          D. Alloin\inst{3}
          \and
          E. Telles\inst{1}
          }

   \institute{Observat\'orio Nacional, Rua General Jos\'e Cristino,
     77, 20921-400, S\~ao Cristov\~ao, Rio de Janeiro,
     Brazil\\ \email{egallian@on.br,etelles@on.br} \and European
     Southern Observatory, Karl-Schwarzschild-Str. 2, 85748 Garching,
     Germany\\ \email{mkissler@eso.org} \and Observatoire de
     Haute-Provence, OAMP, 04870, Saint Michel l'Observatoire,
     France\\ \email{danielle.alloin@oamp.fr}}

   \date{Received ---; accepted ---}
%\abstract{BLABLA}{BLABL}{BLABLA}{BLABLA}{BLABLA} 
% 5 {} token are mandatory
 
  \abstract{This paper presents and examines new near-infrared
    integral field observations of the three so-called 'embedded star
    clusters' located in the nuclear region of
    NGC1365. Adaptive-optics- corrected K-band data cubes were
    obtained with the ESO/VLT instrument SINFONI. The continuum in the
    K-band and emission lines such as \ion{He}{I}, \Bg, and several
    \HH~lines were mapped at an achieved angular resolution of
    0.2\arcsec~over a field of 3$\times$3\,\arcsec$^{2}$ around each
    source. We find that the continuum emission of the sources is
    spatially resolved. This means that they are indeed cluster
    complexes confined to regions of about 50\,pc extension. We
    performed robust measurements of the equivalent width of the CO
    absorption band at 2.3\micro~and of \Bg. For the main mid-infrared
    bright sources, the data only allow us to determine an upper limit
    to the equivalent width of the CO bands. Under the assumption of
    an instantaneously formed standard initial mass function
    Starburst99 model, the new measurements are found to be
    incompatible with previously published mid-infrared line
    ratios. We show that an upper mass limit of 25 to 30\,\msol, lower
    than the typically assumed 100\,\msol, allows one to simply remove
    this inconsistency. For such a model, the measurements are
    consistent with ages in the range of 5.5\,Myr to 6.5\,Myr,
    implying masses in the range from 3 to 10 $\times$
    10$^6$\,\msol. We detect extended gas emission both in \HII~and
    \HH. We argue that the central cluster complexes are the sources
    of excitation for the whole nebulae, through ionisation and shock
    heating. We detect a blue wing on the \Bg~emission profile,
    suggesting the existence of gas outflows centred on the cluster
    complexes. We do not find any evidence for the presence of a lower
    mass cluster population, which would fill up a ``traditional''
    power law cluster mass function.}

   \authorrunning{Galliano et al.} 
   \titlerunning{cluster complexes in NGC1365}

   \maketitle

%----------------------------------------------------------------
%                          INTRODUCTION
%----------------------------------------------------------------
\section{Introduction}
\label{introduction}

The knowledge about young star cluster populations in galaxies
increased much since high angular resolution ($\sim$0.1\arcsec)
visible images could be obtained with the Hubble Space Telescope
\citep[see the reviews by][and references
  therein]{Zwart10,Larsen11}. In the infrared, images approaching such
a high angular resolution can be achieved only from large aperture,
hence ground-based telescopes. The high angular resolution infrared
observations therefore suffer from the limitations inherent to the
technique itself: only a few infrared transparent windows exist, and,
due to the intense atmospheric emission, the sensitivity is relatively
low. The few high angular resolution mid-infrared (MIR) studies of
actively star-forming regions in external galaxies have been able to
resolve them in hot spots, which generally have no counterpart in the
visible, but are bright in the radio. These types of sources have been
named ``embedded clusters''
\citep{Gorjian01,Plante02,Beck02,Martin-Hernandez05,Galliano05a,Martin-Hernandez06,Wold06,Gandhi11}. They
correspond to the hypothetical first phase of a star cluster, when it
is still embedded in its birth material. It is believed that this
embedded phase only lasts for a few million years: the intense stellar
winds and supernova activity provokes the expulsion of the remaining
gas and quickly leads to the disruption of the cluster itself
\citep[see the review by][and references therein]{Bastian11}.

Three examples of massive star clusters in this an evolutionary stage
have been discovered in the star-forming nuclear region of the
strongly barred spiral NGC1365 \citep[][GA05]{Galliano05a}. The
relatively small distance of NGC1365 (18.6\,Mpc, 90\,pc per arcsec),
the fact that this galaxy has been extensively observed and modelled
\citep[see][for a review on the specific galaxy]{Lindblad99}, and the
fact that the galaxy hosts an active galactic nucleus (AGN) that can
feed an adaptive optics (AO) system make these clusters preferred
targets for detailed studies. They are located in the central
star-forming region, close to, or within, a well-defined dust lane. They
were named M4, M5 and M6, 'M' standing for MIR (GA05).

\citet{Galliano05a} presented the discovery (with ESO/3.6m/TIMMI2) of
the bright MIR counterparts to previously known centimetre sources
\citep{Sandqvist82,Saikia94,Sandqvist95,Forbes98}\textbf, and
interpreted them as 'embedded star clusters'.  \citet{Galliano08b},
hereafter GA08, complemented these observations with near-infrared
(NIR) data (K and L images and spectra with ESO/VLT/ISAAC), as well as
new MIR data (narrow 12.8\micro~image and N-band spectra with
ESO/VLT/VISIR). These sources are bright in the MIR continuum, nebular
lines, cm continuum and are spatially associated to CO hot-spots
\citep[][hereafter SA07]{Sakamoto07}.  In GA08, we determined masses
of about several times $10^7$\,\msol~for the stellar component of each
cluster, which puts them in the range of the most massive known
stellar clusters \citep{Kissler06}. The masses were estimated using
the dereddened \Bg~fluxes, and a determination of the cluster
ages. Because the hydrogen emission line fluxes dramatically decrease
with time during the first $\sim$10\,Myr of a cluster life, the mass
estimated using the nebular line flux is very sensitive to the age
determination.  In GA08, we proposed an age determination for the
clusters using the ionised gas emission lines in the N-band. Three
lines of different ionisation potential ([\ArIII] 27.6eV, [\SIV]
34.8eV and [\NeII] 21.6eV) are located in the N-band. In the VISIR
N-band spectra, only the lower ionisation line [\NeII] could be
detected. The stellar population synthesis model for young and
instantaneously formed clusters predicts a drop of more than two
orders of magnitude in the stellar continuum for radiation capable of
ionising [\ArIII] and [\SIV], while this drop is much smaller (one
order of magnitude) for the [\NeII] ionising radiation (See Fig. 15,
16 and 17 in GA08). From this argument, we estimated an age of about
7\,Myr~for the three sources, from which we derived the above
mentioned masses. Some other observational facts supported our age
estimate for the clusters.  First, similar-quality MIR spectra of a
younger massive embedded cluster \citep[NGC5253/C2 with
  ESO/3.6m/TIMMI2 by ][]{Martin-Hernandez05} indeed display intense
[\SIV]. They show that [SIV] is much brighter than [NeII]. Also, the
centimetre spectral indexes of embedded clusters in NGC1365 reveal a
significant share of non-thermal emission, possibly originating from a
high supernovae rate. This points towards ages older than 3\,Myr or
4\,Myr (GA05).  In contrast, younger ($\sim$1\,Myr old) clusters have
flat or even inverted spectra \citep{Kobulnicky99}. The long-slit
K-band ISAAC spectra published in GA08 also displayed the CO
absorption bands at 2.3\micro. These features are only expected to
appear with the first red super giant stars, hence not before
6\,Myr. Nevertheless, since the ISAAC spectra have relatively low
angular resolution, this argument was not considered to be robust
because we could not infirm the possibility that the absorption came
from the background or nearby sources. A more reliable analysis is
presented here, and interestingly, does not confirm the presence of
the CO absorption bands that originates from the clusters themselves.

\citet{Elmegreen09} examined how these sources might have been formed.
These authors showed that the extreme stellar masses derived for the
clusters in GA08 are consistent with the properties of the galaxy: the
large amounts of gas found in the central kpc region that was detected
through HI \citep{Jorsater95} and CO observations \citep{Sakamoto07}
are sufficient for the formation of a few 10$^7$\,\msol~clusters.
\citet{Elmegreen09} propose a scenario to explain the building-up of
the nuclear mass, where gas loops in along dust filaments to feed the
bar dust lane and finally flows inwards towards the nuclear
region. The interaction between the dust filaments and the dust lane
is possibly a trigger for the formation of star clusters, which then
would move inwards following the gas flow and feed the central bulge.

In the present paper, we discuss a new dataset for these three
targets, recorded with SINFONI, the adaptive optics near-infrared
integral field instrument available at the VLT. The observations were
obtained with the adaptive optics loop closed on the nucleus of the
galaxy, and hence have an angular resolution that is close to the
diffraction limit of the 8m aperture telescope. Since the SINFONI data
cubes contain very much information, we give priority to the detailed
presentation of the data, including the publication of tables. We
report about the collection and reduction of the data in
Sec.~\ref{data_collection}. The data analysis including the different
measurement procedures used are described in
Sec.~\ref{data_analysis}. In Sec.~\ref{observational_results}, the
observational results are provided; they are discussed in
Sec.~\ref{discussion}, and our conclusions are given in
Sec.~\ref{conclusion}.

%At the distance of NGC1365 (18.6\,Mpc), the size of the field of view
%around each of the three clusters corresponds to 270\,pc and the
%angular resolution (0.19\arcsec) to 17\,pc.

%----------------------------------------------------------------
%                          DATA COLLECTION
%----------------------------------------------------------------
\section{Data collection and reduction}
\label{data_collection}

The dataset was obtained with SINFONI at the ESO/VLT in the night from
2005/12/04 to 2005/12/05\footnote{program ID: 076.B-0372(A)}. SINFONI
is a NIR integral field spectrometer (SPIFFI) coupled to a visible
curvature AO system. The wavefront was analysed directly on the AGN of
the galaxy (type\,1.5, R=15.4).  The following observing mode was
selected: field of view of 3$\times$3\arcsec$^2$ and spaxels of
0.100$\times$0.050\arcsec$^2$, wavelength range from 1.95\micro~to
2.45\micro~and spectral resolution of 4000, corresponding to an
instrumental broadening of \fwhm=5.5\,\AA~(this translates into
\fwhm=75\kms~or $\sigma$=30\kms). Since the angular distance between
the clusters is about 4\arcsec, we had to point on each target
separately. The total integration time for each pointing was set to
1800\,s. The observations were carried out under regular and stable
atmospheric conditions, with a seeing of 1.2\arcsec, a Strehl ratio of
42$\%$, leading to an angular resolution of about
0.2\arcsec~($\lambda$/D=0.05\arcsec~at 2.1\micro), measured on the
data, as detailed in Sec.~\ref{measurement_procedures}.

An attempt to fully reduce the data cubes using the ESO pipeline
\citep{Modigliani07} was first performed, but left too strong sky line
residuals on the final spectra. These residuals result from the change
of sky line intensity between the observation of the object (OBS) and
that of the sky (SKY). P-Cygni-like profile residuals are also
observed in some cases. They are due to a slight shift in the
wavelength calibration between the OBJ and SKY telescope
positions. These effects are thoroughly discussed in \citet{Davies07}
and specific routines have been developed to correct them.  They were
kindly provided to us by R. Davies and allowed to improve the sky
subtraction. All details about these procedures can be found in
\citet{Davies07}.

We used the following tools in the reduction process: the ESO
pipeline, the routines by R. Davies, our own \texttt{IDL} routines, and the
data visualisation tool \texttt{QFitsView}, developed by Thomas Ott at
MPE\footnote{http://www.mpe.mpg.de/$\sim$ott/QFitsView}.

The observation of each target was split into two blocks (two OBs),
each of which is a sequence of six integrations alternating object and
sky (OBJ-SKY-SKY-OBJ-OBJ-SKY). For each field, we hence had six pairs
(OBJ+SKY) of cubes. A standard jittering procedure was used for each
sequence.

\begin{figure}[htbp]
\begin{center}
\includegraphics[width=8cm]{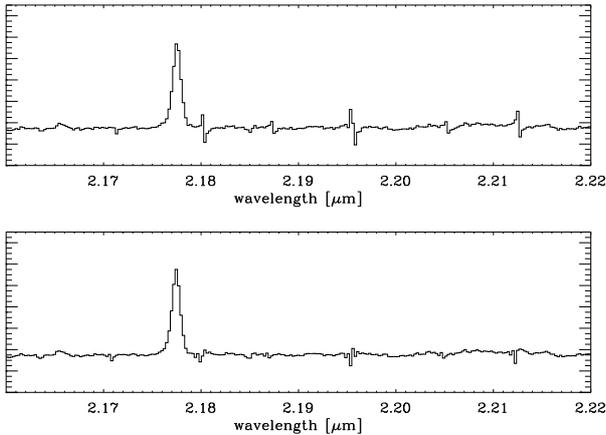}
\caption{Example of the sky-subtraction correction, for an average
  case. We display a portion of the 1\arcsec~radius aperture spectrum
  of one of the M4 data cube pairs. On top, the P-Cygni-like residuals
  are stronger before correction, and below, after correction, the
  residuals have been attenuated. The emission line is \Bg.}
\label{skysub}
\end{center}
\end{figure}

The data reduction was performed as follows. First, the raw frames
were cleaned from the bad lines that are created by the data
processing hardcoded at the detector level. The correction was
performed using the routine provided in the SINFONI data reduction
cookbook. We then reduced the calibration frames using the
pipeline. Second, pipeline basic reduction was applied to each
individual science cube (and not to each individual pair of cubes, as
would be the case for standard pipeline reduction): the data were
flat-fielded and corrected for dead/hot pixels; interpolations to
linear wavelength and spatial scales were applied, after which the
slitlets were aligned and stacked to create the final cube. The
reduced sky cubes were then subtracted from the corresponding object
cube using the routines from \citet{Davies07}. A sample of a spectrum
without/with residual correction is displayed in Fig.~\ref{skysub}. It
shows that we did not entirely get rid of the artefact P-cygni like
residuals. Hence, for enhancing the legibility of the data, we decided
to perform a cosmetic step after the correction, that replaced the
pixel values by the median value over the wavelength range of the main
lines. These ranges are highlighted in grey on the spectra displayed
in the paper. The only line of interest that is critically affected
is \HH~3-2\,S(3).

%----------------------------------------------------------------------
%----------------------------------------------------------------------

Then, for each pointing, the six sky-subtracted reconstructed cubes
were combined. For this, relative astrometry was performed
between the cubes, using the header information about the jitter
sequences.

We simultaneously performed the relative wavelength response and
telluric feature corrections using the ESO provided data products of
27 observations of telluric standard stars. A median response,
corrected from the stellar features, was then computed using the
stellar spectra published by \citet{Pickles98} that are available on
the ESO website. A scaling factor, following the \texttt{Iraf
  telluric} routine, was applied to adjust the depth of the
atmospheric features. We also used this set of standard star data to
evaluate the uncertainty on the flux calibration, which we call
hereafter ``photometric uncertainty'', in contrast to ``measurement
uncertainty''. The standard deviation of the conversion factors for
these star flux measurements is 40\%. Because these observations were
obtained on different nights, we can be confident that the
``photometric calibration uncertainty'', due to both the uncertainty
on the standard star fluxes and the sky transparency variations, is at
most $\pm 20$\%.

To perform the relative registration of the cubes, we first built
K-band images by convolving by the filter transmission, and then
degraded the image and pixel size to adjust it to that of the ISAAC
K-band image published by GA08. For each image, we ascertained the
world coordinates of the brightest peak using the ISAAC image to
register the cubes. We hence used the same astrometric reference as in
GA08, where the [0\arcsec;0\arcsec] coordinate was set to the peak of
the AGN in the VISIR 12.8\micro~image.

Once the registration was performed, we use the comparison between the
degraded SINFONI K-band reconstructed images and the ISAAC images to
perform the absolute flux calibration of the images. This choice
implies that the flux measurements on the ISAAC and on the SINFONI
data agree by definition.

The last step of the reduction is the correction of the wavelength
calibration to set the radial velocities in the local standard of rest
(LSR).

%----------------------------------------------------------------
%                        DATA ANALYSIS
%----------------------------------------------------------------

\section{Data analysis}
\label{data_analysis}

This section presents the methods used to extract the quantitative
information from the dataset.

\subsection{Map reconstruction}
\label{map_reconstruction}

%----------------------------------------------------------------------
\begin{figure*}[htbp]
\begin{center}
\includegraphics[width=17cm]{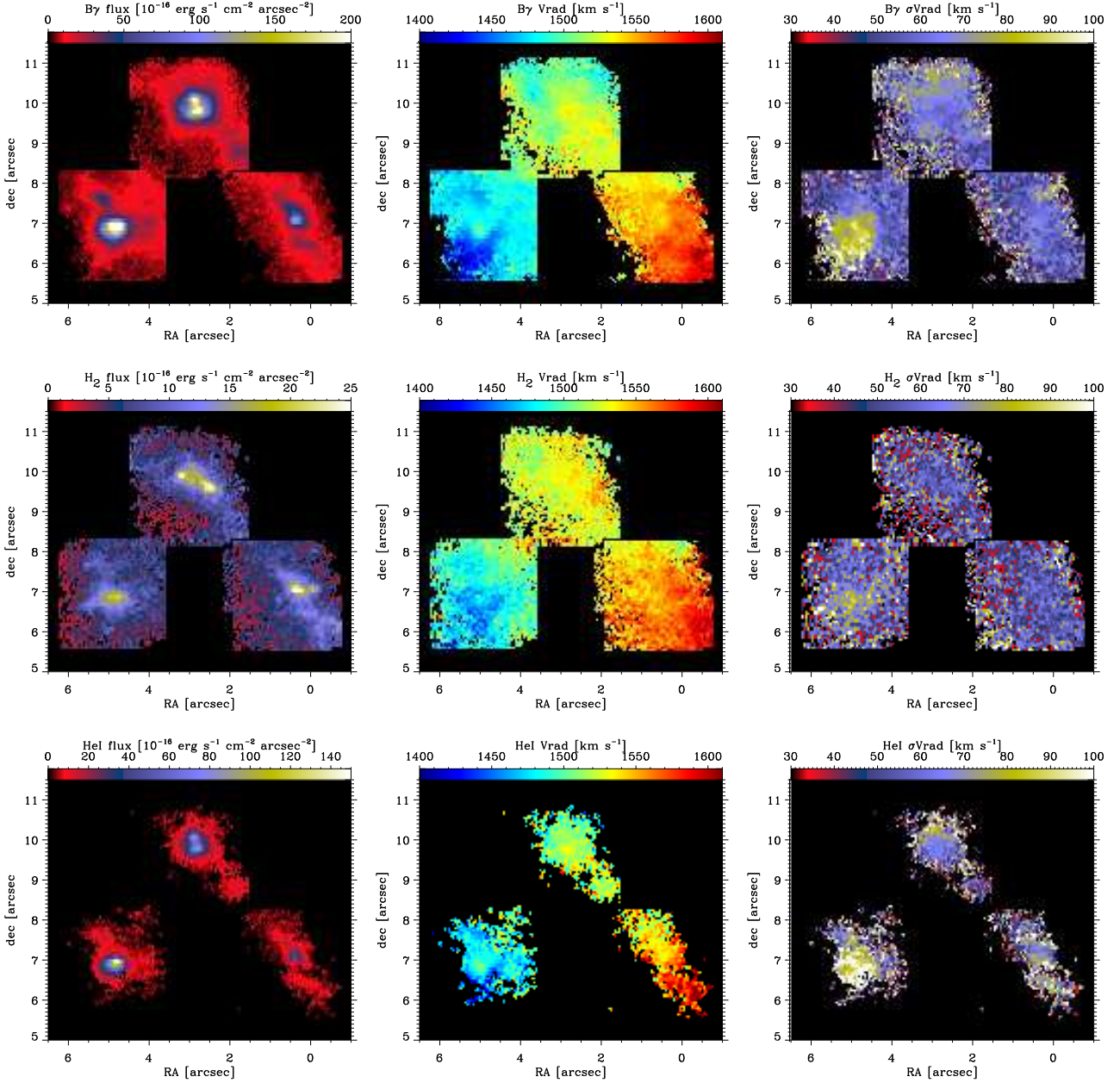}
\caption{Gas emission around the extremely massive clusters in
  NGC1365, observed with VLT/SINFONI. First row: \Bg~flux, radial
  velocity, and velocity dispersion, from left to right. Second row,
  same for H2; third row, same for \ion{He}{I}. The axes are in
  arcsec, north is up and east is left.}
\label{study_maps_big_lines}
\end{center}
\end{figure*} 
%----------------------------------------------------------------------
\begin{figure*}[htbp]
\begin{center}
\includegraphics[width=18cm]{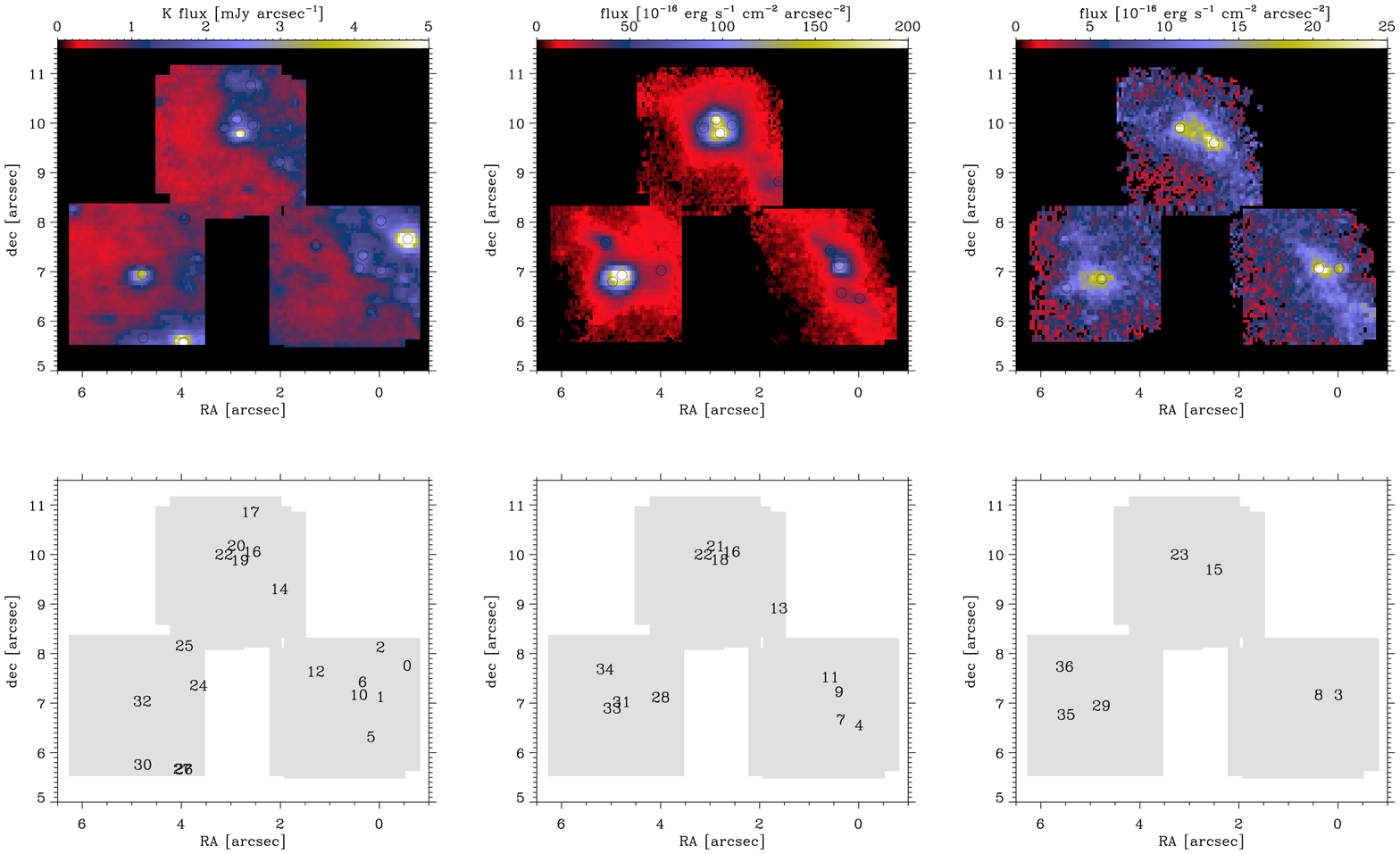}
\caption{Top row, left to right: reconstructed maps for K-band flux
  density, \Bg~flux and \HH~1-0\,S(1)~flux. The chosen apertures are
  drawn on the maps. The aperture labels are given for each map on the
  empty map below it for clarity. The label number is given in increasing Right
  Ascension. North is up and east is left, and the X and Y axes are in
  arcsec.}
\label{apertures}
\end{center}
\end{figure*} 
%----------------------------------------------------------------------

The first step of the analysis of the reduced data cubes is
reconstructing 2-D maps from 3-D cubes. For three chosen emission
lines (\ion{He}{I} 2.06\micro, \HH~2.12\micro~and \Bg~2.16\micro), we
performed an automatic Gaussian fit for each spectrum of the cubes,
and built 1) the radial velocity maps (Gaussian profile central
wavelength), 2) the radial velocity dispersion maps ($\sigma$ of the
Gaussian), and 3) the Gaussian flux maps from the fit parameter
values. These maps are displayed in Fig.~\ref{study_maps_big_lines}:
from left to right, line flux, line radial velocity, and line width. A
K-band map is obtained by convolving each spectrum by the filter
transmission curve. A narrow band (excluding any spectral feature) map
is also computed; it shows that the difference to the full K-band map
is negligible. Accordingly, we only considered the full K-band maps. The
K-band reconstructed map is displayed in Fig.~\ref{apertures} in the
top left frame. This figure will be described in
Sec.~\ref{measurement_procedures}, when we define the set of apertures
used for performing the measurements.

The \Bg~large-scale kinematics (radial velocities) are consistent with
the CO kinematics. The \Bg~velocities can be compared to the
large-scale isovelocity contours in SA07, which clearly show a
rotation pattern with non-circular motions. From their table\,5, the
measured LSR velocities (at the peak of brightness temperature for
$^{12}$CO(2-1)\,) are 1560\kms, 1540\kms and 1480\kms~for the CO knots
associated with M4, M5 and M6. In the SINFONI data, for a
1.0\arcsec~diameter aperture, we find 1555\kms, 1520\kms and
1475\kms~for \Bg, and 1560\kms, 1530\kms, and 1470\kms~for
\HH~respectively. We measured these velocities with a simple Gaussian
fit at the positions [0.41;7.12] , [2.80;9.86] , [4.80;6.90]. These
results indicate that the \Bg~velocities are slightly blueshifted with
respect to CO. The $^{12}$CO(2-1) \fwhm~are 70\,\kms, 60\,\kms~and
80\,\kms~respectively, while \Bg~is broader with an \fwhm~of 145\kms,
145\kms~and 200\kms~respectively.

\subsection{Measurement procedures}
\label{measurement_procedures}

To perform measurements on the data cubes, we chose to first define a
set of significant apertures to reduce the amount of information to be
analysed. We used the K-band, \Bg, and \HH~reconstructed maps to
select the positions of these apertures. We chose the most prominent
sources in the three images, as well as a few other apertures that
sample regions of lower surface brightness. These apertures are drawn
and labelled over the reconstructed maps in Fig.~\ref{apertures},
which show from left to right, the K-band, \Bg~and \HH~maps. Below
each map we show the set of labels used for identifying the apertures:
a number increasing with Right Ascension. Their radius of
0.1\arcsec~allows one to sample single angular resolution
elements. The measurements performed are consequently the least
possibly affected by source mixing. Results for most measurements
described below are given in table~\ref{table1}, \ref{table2},
\ref{table3} and~\ref{table4} and can be used for subsequent
analysis. The line fluxes given in the tables have not been corrected
for any extinction.

The uncertainties on the different measurements have two origins: the
uncertainty due to the noise and the measurement procedure, and the
uncertainty on the calibration of the data, which we called
photometric uncertainty in Sec.~\ref{data_collection}. In the
tables, we only give the measurement uncertainties. The total maximum
uncertainties for the fluxes can be obtained by quadratically adding
the $\pm$20\%~uncertainty derived in Sec.~\ref{data_collection} to the
measurement uncertainty.

%\subsubsection{Measurement of the angular resolution}
%\label{angular_resolution}

%----------------------------------------------------------------------
\begin{figure}[htbp]
\begin{center}
\includegraphics[width=9cm]{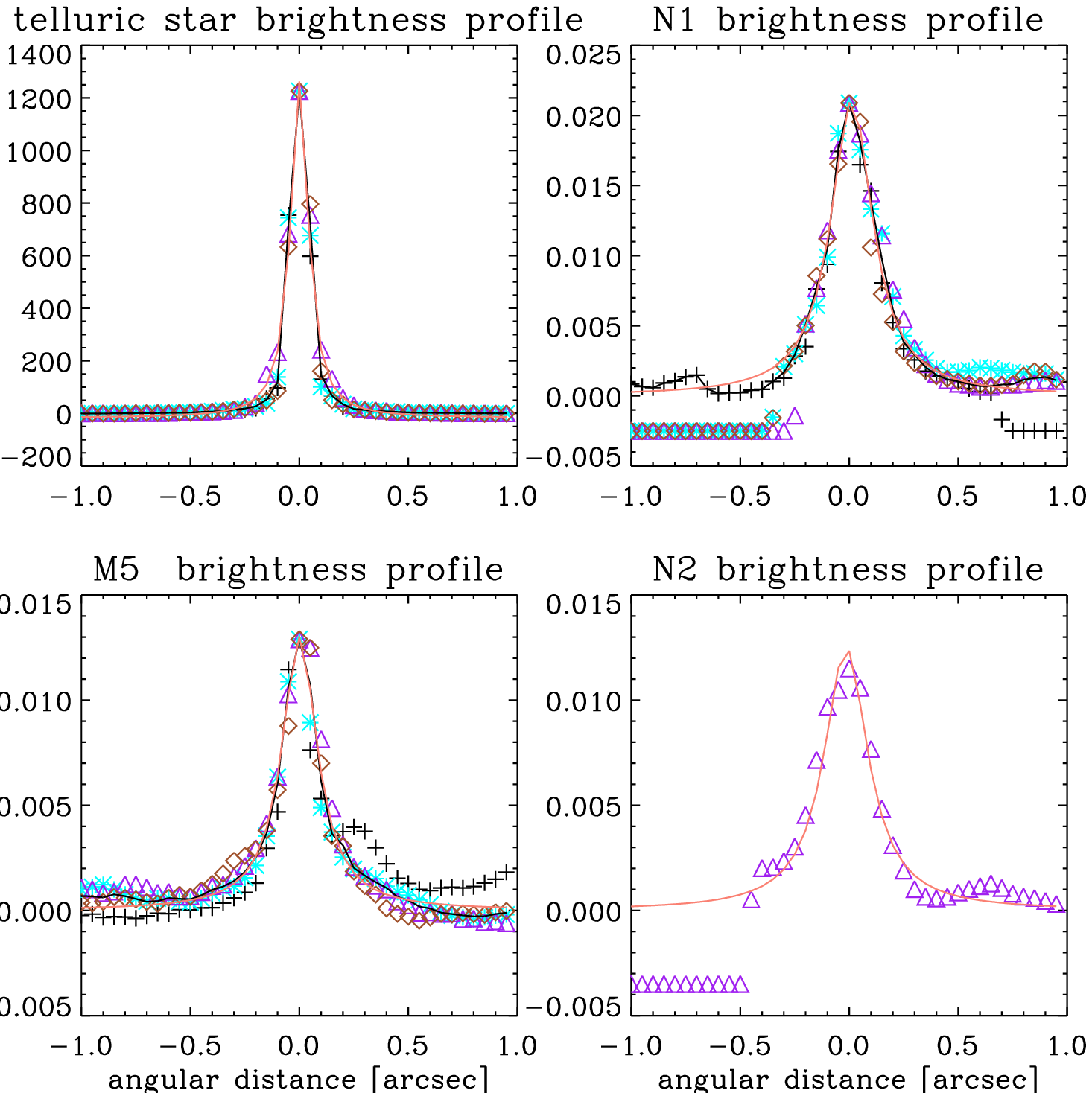}
\caption{Light profiles (1D cuts through the images) for a telluric
  star, N1, M5 and N2. For the first three, we took cuts along
  PA=0\degr,45\degr,90\degr~and 135\degr (black crosses, cyan
  asterisk, purple triangle and brown diamond). The median profile is
  drawn with a black line, and the Lorentzian fit with a salmon
  line. For N2, we only have PA=0\degr~since it is on the edge of the
  image. The fit \fwhm~are 0.10\arcsec, 0.23\arcsec,
  0.20\arcsec~0.25\arcsec.}
\label{spat_prof}
\end{center}
\end{figure} 
%----------------------------------------------------------------------

\begin{figure}[htbp]
\begin{center}
\includegraphics[width=9cm]{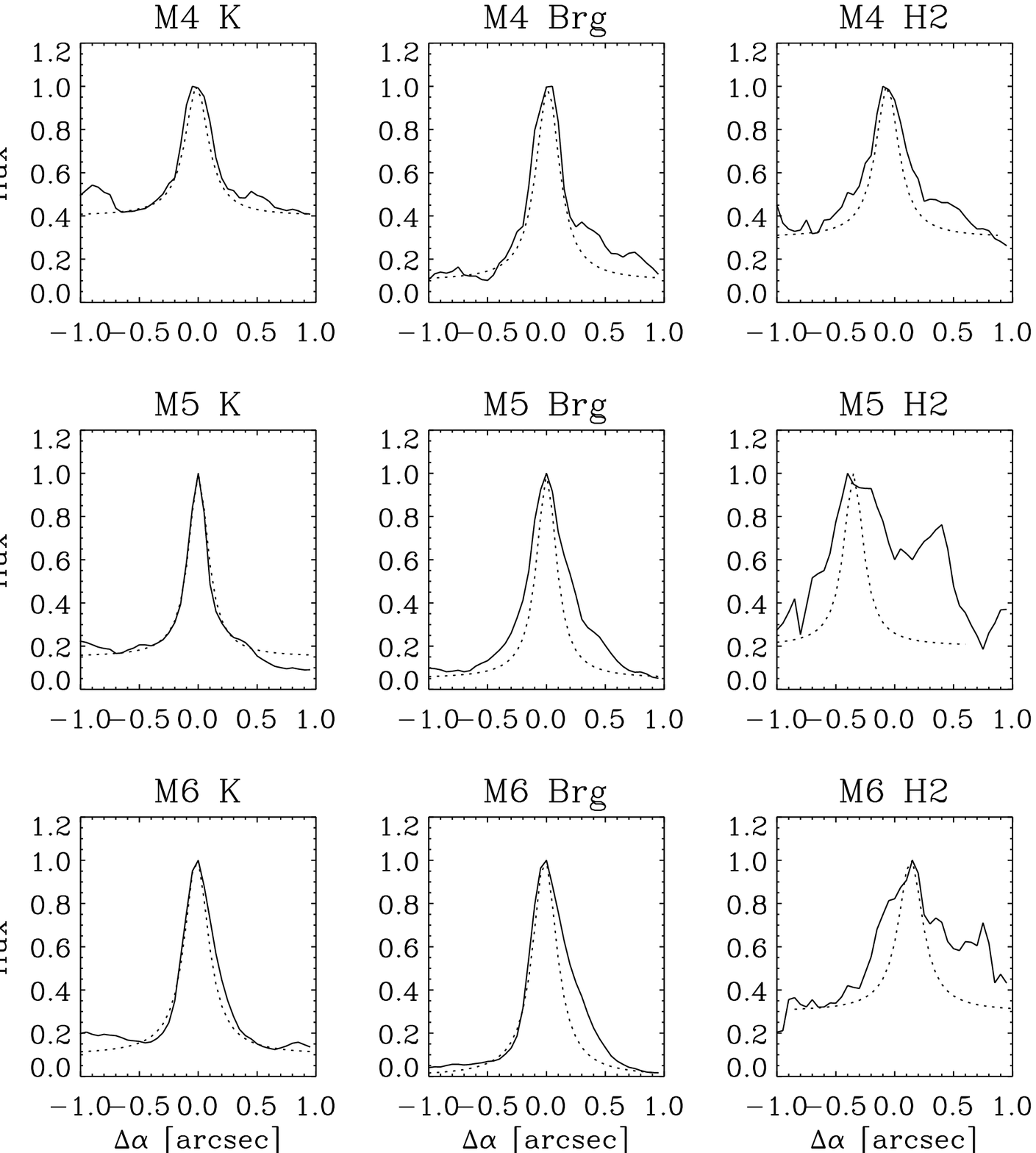}
\caption{Profile cuts for M4, M5 and M6 (position given is the table
  for the light centre on the K-band image) along PA=45\degr, 60\degr,
  and 110\degr. The dotted line is the comparison profile for each
  cube (as explained in the corresponding paragraph). It has been
  rescaled to give a meaningful comparison.}
\label{bright_dist}
\end{center}
\end{figure}

To measure the angular resolution on each cube, we chose sources that
are bright and narrow and measured their profiles on the K-band
reconstructed image. We chose four sources: a telluric star, the
K-band bright cluster N1 (to the west of M4), M5 and N2, the bright
cluster on the southern part of the M6 field. The profile measurements
are shown in Fig.~\ref{spat_prof}. The cuts are obtained along
PA=0\degr, 45\degr, 90\degr~and 135\degr, at the position of the
source given in table~\ref{table1}, and at the centre of light for the
star. Since the data are AO-corrected, it is natural for the telluric
star to have a narrower profile, because the Strehl ratio is
higher. In the best case, the achieved resolution would hence be
0.1\arcsec~\fwhm. For our dataset, the values measured are
0.23\arcsec, 0.20\arcsec~0.25\arcsec, for the M4, M5 and M6 cubes,
respectively, and are to be considered as upper limits, since we cannot be certain that the sources are not resolved. We used the fits of
the spatial profiles for comparison with the profiles measured on the
data. In Fig.~\ref{bright_dist}, we compare the profiles of the main
sources (M4, M5 and M6) for K-band, \Bg~and \HH~2-1\,S(1) to the
measured reference profiles to check if they are resolved.

%----------------------------------------------------------------------
%----------------------------------------------------------------------
\begin{figure*}[htbp]
\begin{center}
\includegraphics[width=18cm]{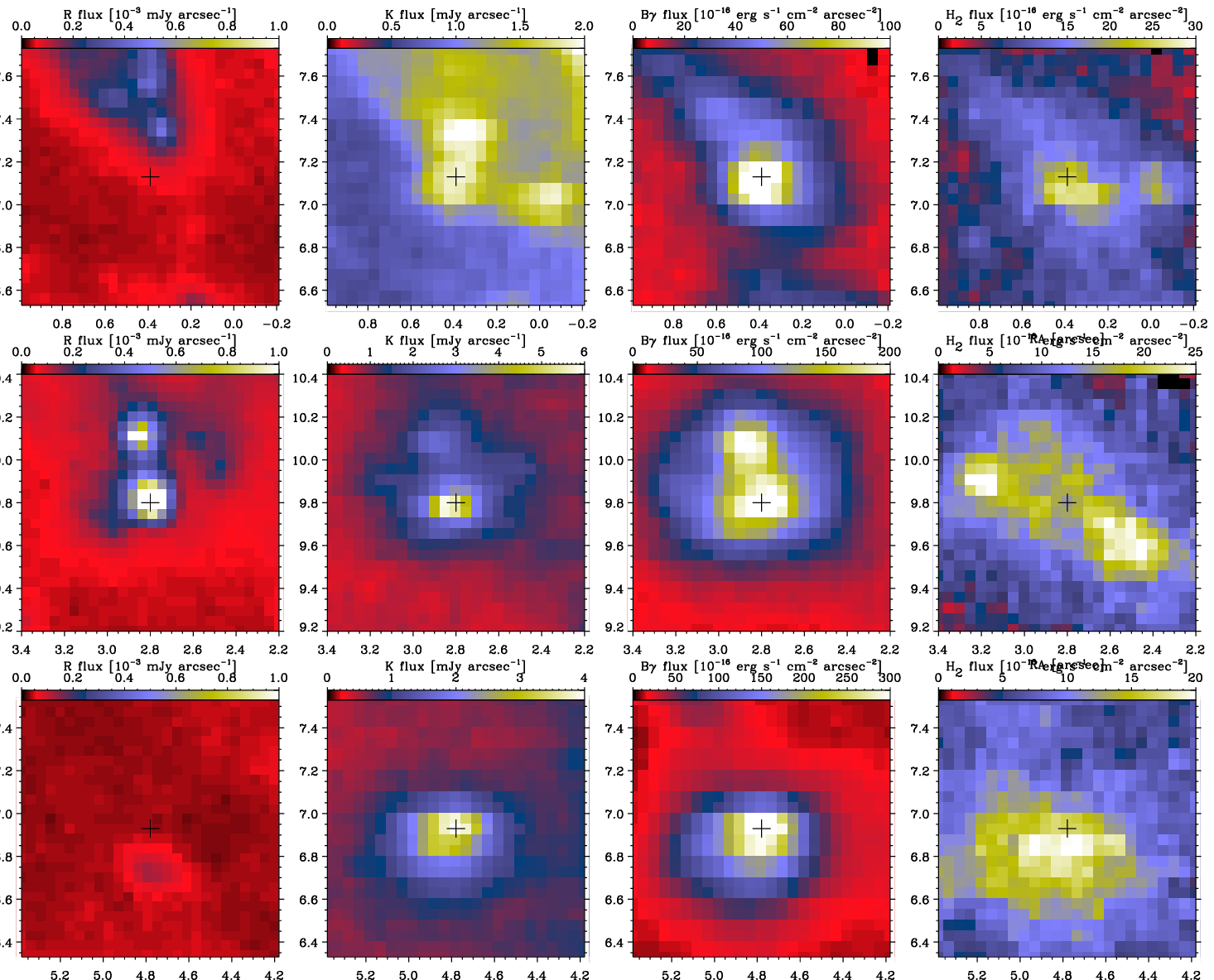}
\caption{Zooms on the 1.2$\times$1.2\arcsec$^2$ around M4 (first row),
  M5 (second row) and M6 (third row). For each source, from left to
  right we display the R-band image (HST), the K-band image emission,
  the \Bg~emission and the \HH~1-0S(1) emission. The crosses mark the
  position of the \Bg~peak. North is up and east is left, and the X
  and Y axis are in arcsec.}
\label{study_maps_details}
\end{center}
\end{figure*} 
%----------------------------------------------------------------------

%\subsubsection{Definition of the set of apertures} 
%\label{source_definition}

The centre of each aperture was first chosen by eye on the
reconstructed maps, and then taken as the centre of light in a box of
3$\times$3 pixels around the initially chosen position. For apertures
that coincide with emission peaks, the given aperture position is a
measurement of the source position. For apertures defined on extended
emission regions, the precise position is not meaningful. Different
peak positions coming from measurements inside a single cube (for
example the K-band peak and a \Bg~peak) have a good relative
registration precision, up to a fraction of pixel size (1/5 of a pixel
corresponds to 0.01\arcsec). Between cubes, the registration was
performed using the ISAAC images, and hence has the precision of a
fraction of an ISAAC pixel size (1/5 corresponds to 0.03\arcsec).

%\subsubsection{Measurement of the line fluxes and widths}
%\label{line_flux}

The emission line fluxes through the given apertures are derived from
automatic Gaussian+continuum fitting of the line profiles, carried out
with the \texttt{idl} routine \texttt{mpfitpeak} by
\citet{Markwardt09}. We measured the following lines: \Bd, \Bg,
\ion{He}{I}, and \HH~roto-vibrational 1-0\,S(3), 1-0\,S(2), 2-1\,S(3),
1-0\,S(1), 2-1\,S(2), 1-0\,S(0), and 2-1\,S(1). The errors on the
fluxes and widths were derived from the routine results, whereas the
error on the line centre was simply taken to be 1/5 of the line
\fwhm. For the flux measurements for lines below 2\micro, the
uncertainty is greater due to the uncertainty on the response, because
of the poorer atmospheric transmission in this region. We evaluated the
uncertainty using the spectrum of M6 (apertures 31 and 32), which
displays a steep continuum. If we extrapolate the continuum slope down
to 1.95\micro, we find 0.0015\,mJy, while we measure 0.0025\,mJy on
the spectrum. This means that the error can be higher than 50\%.

%\subsubsection{Measurement of the equivalent width of the CO 2.29\micro~absorption band}
%\label{EW(CO)}

We measured the equivalent width of the CO 2.29\micro~absorption band
(EW(CO)) using the following procedure. First, we built a spectrum of
the background emission of the region. Using a histogram of the K-band
flux density, we assumed that the pixels with less than
0.7mJy\,arcsec$^{-2}$ belong to the background. A median background
spectrum was then built for each of the three cubes. For each
aperture, we measured EW(CO) on the background-subtracted
spectrum. The continuum fit was performed over three intervals between
the bands, while the depth of the feature is measured between
23050\,\AA~and 23160\,\AA. The measurement error given in
table~\ref{table1} corresponds to $\pm$ 0.3$\sigma$ of the continuum
around the fit. If the continuum signal to noise ratio (S/N) was less
than 3, we only considered the upper limit of the EW(CO) (feature
width divided by 0.3$\sigma$).

%\subsubsection{Measurement of the K-band and \Bg~fluxes on the reconstructed images}
%\label{image_fluxes}

At the defined aperture positions, we performed larger aperture
background-subtracted flux measurements (aperture radius of
0.2\arcsec~and background in an annulus between 0.2 and
0.6\arcsec). We used the \texttt{djs\_phot} for the measurement and
error evaluation. From these two measurements, we extract what we
refer to in the following as the equivalent width of \Bg: EW(\Bg).

%\subsubsection{Measurement of the continuum slopes}
%\label{continuum slope}

We measured the slope of the continuum (after background subtraction)
as the ratio between the fluxes measured through two artificial
rectangle 200\,\AA~broad filters, which do not include any spectral
feature and are centred at 20300\,\AA~and 22100\,\AA.  For these flux
densities, the error was set to half the dispersion of the continuum.

%----------------------------------------------------------------
%                     OBSERVATIONAL RESULTS
%----------------------------------------------------------------

\section{Observational results}
\label{observational_results}

This section describes the main observational results derived from the
analysis of the data described above. These results are discussed
in Sec.~\ref{discussion}. First, we present the results related to the
images reconstructed from the data cubes and then those derived from
the spectral measurements.

%-------------------------
%         IMAGES
%-------------------------

\subsection{Images}
\label{images}

Fig.~\ref{study_maps_details} shows the zoomed
1.2$\times$1.2\arcsec$^2$ fields around M4 (top row), M5 (middle row)
and M6 (bottom row). From left to right we display (1) the HST
``R-band'' image (filter F606W), (2) the SINFONI K-band image, (3) the
\Bg~image and (4) the \HH~1-0\,S(1) image. The crosses mark the
position of the \Bg~peak as defined by the procedure described in
Sec.~\ref{measurement_procedures}. It is important to keep in mind
that the pixels on the cubes before combination are rectangular
spaxels, which appear as 0.05$\times$0.1\arcsec$^2$ rectangular
vertical pixels on the reconstructed images. This can give the
illusion that some sources are extended in the north-south direction.

The cluster positions were defined in GA08 as the peak of [\NeII]
12.8\micro~emission. Now, we can refine this positioning by taking the
\Bg~peak position on the 0.2\arcsec~resolution images. The [\NeII] and
the \Bg~positions are consistent, and the difference for M5 arises
because, at the resolution of SINFONI the source is resolved into two
bright knots. The new positions for the clusters are
[0.39\arcsec;7.13\arcsec] for M4/source 10, [2.80\arcsec;9.80\arcsec]
for M5/source 18 and [4.78\arcsec;6.93\arcsec] for M6/source 31. The
absolute precision of this registration with respect to the AGN is
0.03\arcsec~($\pm 0.015$\arcsec, see
Sec.~\ref{measurement_procedures}).

At the new achieved angular resolution, we see that the \Bg~and
\HH~sources that seemed coincident in the former observations at lower
resolution are in fact spatially distinct emission knots. It is
expected that the formerly superposed \Bg~and \HH~knots are not
anymore perfectly coincident when the angular resolution is
sufficiently high. This has previously been observed with a similar
technique in IIZw40 by \citet{Vanzi08}. In our case, the data show
that M4 is marginally resolved in K and \Bg, while the associated
\HH~peak is resolved (\fwhm$\sim$0.3\arcsec) and is shifted by about
0.1\arcsec~to W-SW. A secondary \HH~source (source 3) is detected
about 0.5\arcsec~to the east of M4, and has no counterpart in any
other emission. M5 is clearly resolved into four sources (apertures
16, 18, 21, 22) following a cloverleaf configuration inside a circle
of about 0.5\arcsec diameter. Apertures 18 and 21 correspond to an
emission peak in K and \Bg. These sources are not resolved
individually. The associated \HH~emission lies outside this
cloverleaf, in two resolved knots aligned along PA$\sim$70\degr~and at
a distance of about 0.8\arcsec~(apertures 23 and 15). M6 is detected
and resolved in K and \Bg. It has an asymmetrical shape, with an
extension towards the S-E, which can be seen as an extended wing in
the profiles in Fig.~\ref{spat_prof}. The source as a whole is
detected in \HH, but the maximum of \HH~is displaced by about
0.1\arcsec~to the south with respect to the \Bg~emission.

\medskip

Fig.~\ref{study_maps_big_lines} shows the full combined gas emission
reconstructed maps for \Bg~(top row), \HH~1-0\,S(1) (middle row) and \ion{He}{I}
(bottom row). From left to right we display the line flux, line
centre and $\sigma$ maps. We detect extended emission
on scales of several arcsec, both in \Bg~and \HH. They have the same
global distribution, but differ at the sub-arcsec scale: M4 and M5 are
located on a filament-like emitting structure oriented along
PA$\sim$45\degr~and extending over about 7\arcsec, with a width of
about 1\arcsec. A distinct and more symmetrical emission region of
about 3\arcsec~diameter is observed around M6. The bright
emission peaks are localised well in the middle of the extended
emission regions. Apart from M4, M5 and M6, and their closely associated
sources, there are only few secondary weak peaks (apertures 4, 7, 13 and
28 for \Bg).

%-------------------------
%        SPECTRA
%-------------------------

\subsection{Spectra}
\label{spectra}

%----------------------------------------------------------------------
\begin{figure*}[htbp]
\begin{center}
\includegraphics[width=18cm]{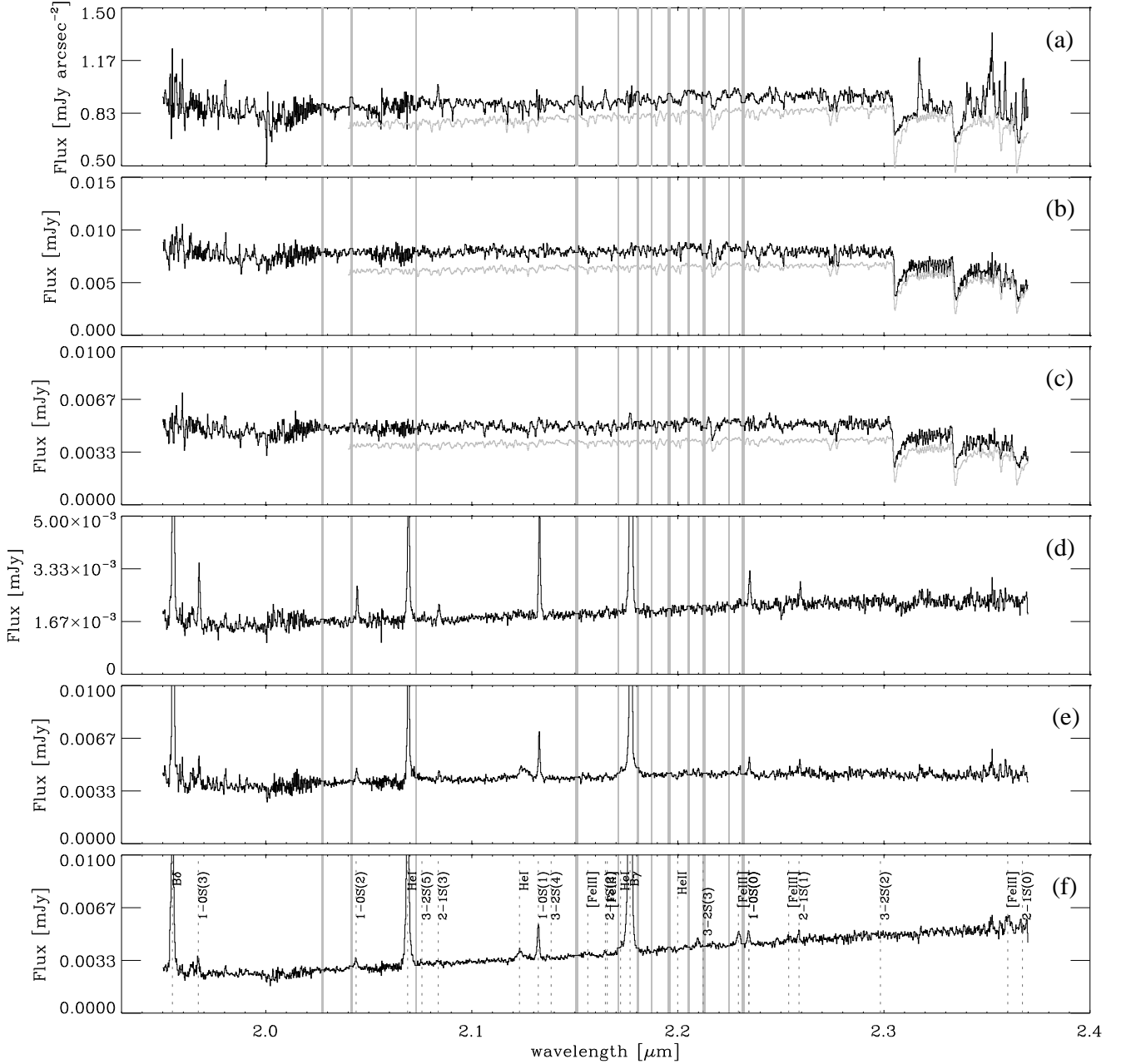}
\caption{Top to bottom: (a) spectra of the background (defined in
  Sec.~\ref{measurement_procedures}) + HR6705 in grey, (b) aperture 0
  + HR2508 in grey, (c) aperture 26 + HR2508 in grey, and (d) aperture
  9 (M4), (e) aperture 18 (M5) and (f) aperture 31 (M6). (d), (e) and
  (f) are not background subtracted.}
  
\label{full_spectra}
\end{center}
\end{figure*} 
%----------------------------------------------------------------------

Fig.~\ref{full_spectra} shows several full-wavelength range spectra of
(a) the background as defined in~Sect.~\ref{measurement_procedures},
(b and c) the two brightest K-band clusters, and (d, e and f) M4, M5
and M6. Over the three first spectra, we plotted in grey normal
star spectra taken from the library of \citet{Wallace97} for
comparison. The line wavelengths given in the following are the rest
wavelengths, the observed redshifted ones are about
0.01\,\micro~longer.

The full-field combined K-band map is displayed on the left of
Fig.~\ref{apertures}. What we defined as background continuum in
Sect.~\ref{measurement_procedures} appears in red (up to
0.7\,mJy\,arcsec$^{-2}$). The corresponding spectrum is given in
Fig.~\ref{full_spectra}.a. Across this background, we mostly detect
continuum sources along the border of the dust lane, on its western
side. Only a few sources are detected on the dust lane itself, except
for M6: we identified the sources corresponding to apertures 5, 12, 24
and 25. Two bright K-band clusters are observed: apertures 0 and
26. Their spectra are given in Fig.~\ref{full_spectra}b,c.

The emission lines initially present on the background spectrum were
removed artificially, and a pure continuum spectrum is left. In grey
we overplot the spectrum of HR6705 (K5III). The next two spectra
(apertures 0 and 26) correspond to the two brightest K-band
sources. From these spectra, no emission lines were removed. The
overplotted stellar templates are HR2508 (M1 + Ib-IIa). The spectral
slopes of the templates were slightly reddened to adjust them to the
observed ones.  The spectra display strong CO absorption as well as
metallic line absorptions. The next three spectra are M4, M5 and M6
(apertures 9, 18 and 31). The three sources show a rising continuum
towards the red and several emission lines, labelled on the spectrum
of M6: \Bd, \Bg, the 2.06 \ion{He}{I}\,\micro~line and seven
\HH~roto-vibrational lines. For M5 and M6, a broad \ion{He}{I}
2.11\micro~observed close to the blue side of \HH~1-0\,S(1) is
detected. The Gaussian fit \fwhm~is 50\,\AA~and 38\,\AA~for M5 and M6,
respectively, while it is about 10\,\AA~for the other lines. Weak
\ion{He}{II} absorption at 2.19\micro~is detected for M5 and M6. For
M6 only, we detect two weak [\ion{Fe}{III}] emission lines at
2.22\micro~and 2.24\micro. An unidentified emission line is detected
on the spectrum of M6 at an observed wavelength of 22097\,\AA, which
corresponds to a rest wavelength of 21990\,\AA.

\medskip

%----------------------------------------------------------------------
\begin{figure}[htbp]
\begin{center}
\includegraphics[width=9cm]{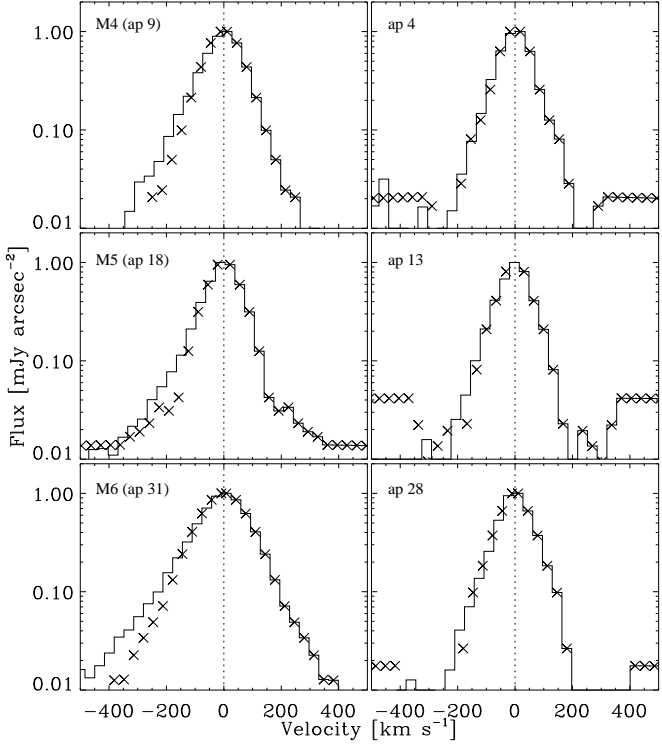}
\caption{Shape of the \Bg~line. Top to bottom, left to right:
  apertures 9 (M4), 18 (M5), 31 (M6), 4, 13 and 28. The null velocity
  is chosen as the centre of the Gaussian fit of the line, and the
  continuum derived in the fit is subtracted. The crosses show the
  actual profile values for the positive velocities, which are also
  plotted for the negative velocities to show the asymmetry of the
  profile.}
\label{line_shapes}
\end{center}
\end{figure} 
%----------------------------------------------------------------------

For M4, M5 and M6, we detect an asymmetry in the \Bg~profile. The same
asymmetry is observed for \ion{He}{I}
2.06\,\micro. Fig.~\ref{line_shapes} displays the profiles for M4, M5
and M6 in the left column from top to bottom, and for a set of
off-centred apertures (4, 13 and 28) in the right column. The
intensities are shown on a logarithmic scale. The null velocity is
computed as the central velocity of the Gaussian fit of the line. For
the positive velocities, the crosses just show the data, and these are
mirrored on the side of the negative velocities, from which one can
appreciate the degree of asymmetry of the profile. Clearly, M4, M5 and
M6 have an extended blue wing, while the spectra from the three other
apertures display symmetrical profiles.

\medskip

\begin{figure*}[htbp]
\begin{center}
\includegraphics[width=18cm]{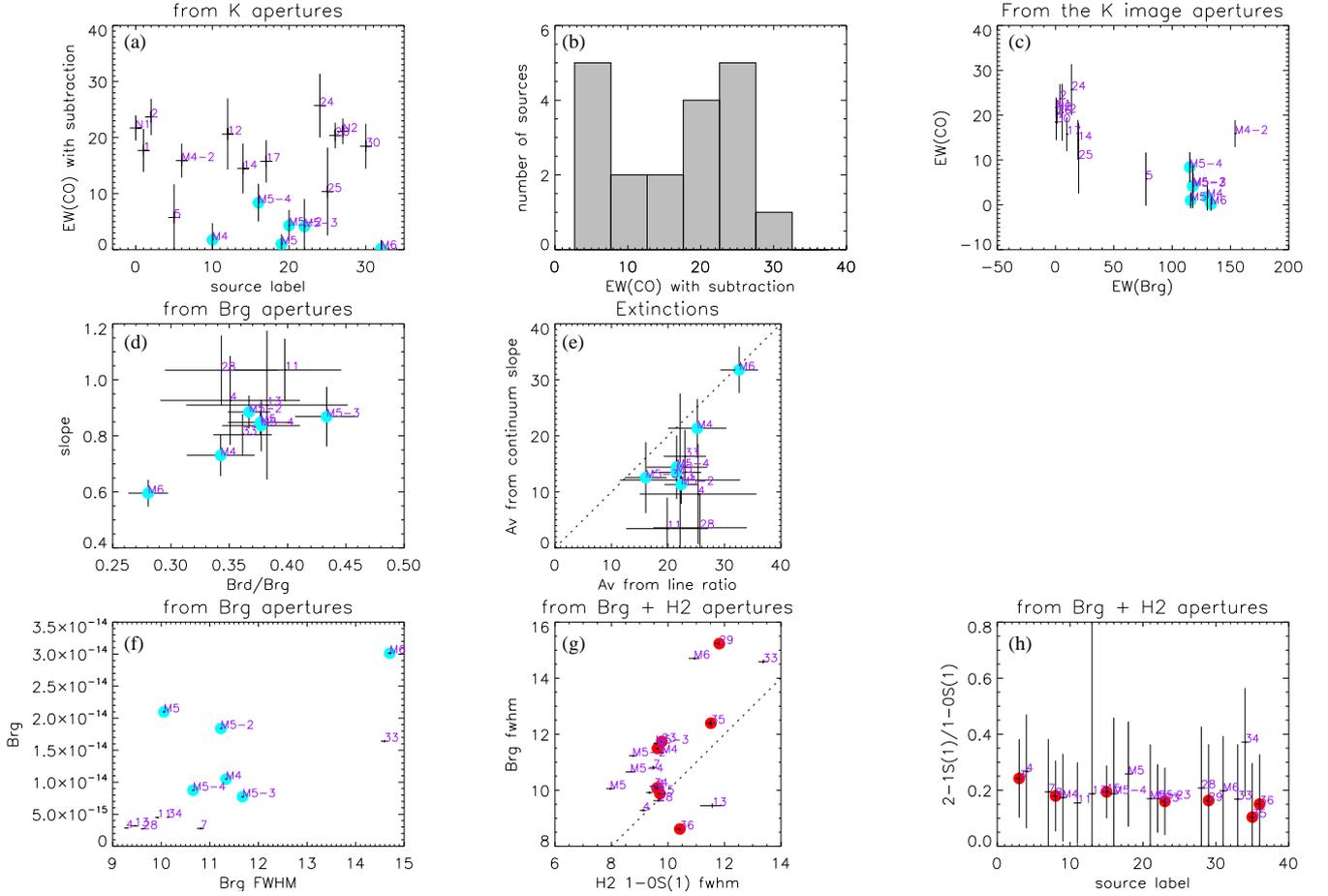}
\caption{Set of plots showing different relations between
  observables. The plots are described in Sec.~\ref{spectra}. In all
  plots, the cyan dots correspond to the sources M4, M5, M5-2 to M5-4
  and M6. The red dots correspond to the apertures chosen in the \HH~image.}
\label{all_plots}
\end{center}
\end{figure*}

Fig.~\ref{all_plots}~gathers a set of diagrams that plot various
observables measured on the data following the procedures detailed in
the former section. For all diagrams, the blue dots mark the bright
\Bg~sources, and the red dots mark the \HH~emission knots. The purple
labels refer to either the aperture numbers as defined in
Fig.~\ref{apertures} or directly the source names for M4, M5 and M6.
As detailed in Sec.~\ref{measurement_procedures}, the equivalent width
of the CO 2.29\micro~band was measured through all defined
apertures. The EW(CO) vs. the aperture number is plotted in
Fig.~\ref{all_plots}.a. Absorption is detected through all apertures,
except for the \Bg~peak sources M4, M5 and M6, for which the upper
limit is a few \AA. For the sources that are not bright in \Bg, the
EW(CO) is about 20$\pm$5\,\AA.  Fig.~\ref{all_plots}.b shows as a
histogram the number of apertures per EW(CO) bin and identifies two
groups. In Fig.~\ref{all_plots}.c, we plot the EW(CO) vs. the
EW(\Bg). We see that the two groups are well identified in both
quantities, with a well-defined anti-correlation between EW(\Bg) and
EW(CO). M4-2 (or aperture 6, the K source right next to M4) has a high
EW(CO), and its also high EW(\Bg) is likely due to its being
superimposed over the extended \Bg~emission from M4. The EW(\Bg) of
the older sources range from 0\,\AA~to 20\,\AA, and their EW(CO) from
15\,\AA~to 25\,\AA. The other sources have higher EW(CO), ranging from
110\,\AA~to 150\,\AA, and low EW(\Bg).

\medskip

From the line analysis of our dataset, the only possibility to
evaluate the extinction is to use the ratio between the two hydrogen
lines that we can measure: \Bd~and \Bg. Unfortunately, as stated in
Sec.~\ref{measurement_procedures}, the uncertainty on the flux
calibration in the region of \Bd~is greater than in the other regions
of the spectral range. Indeed, if we blindly measure the \Bd~fluxes,
we find the \Bd/\Bg~ratio to be higher than that of the theoretical
case B (0.66). We decided to apply a correction factor of
0.0015/0.0025=0.6 to the \Bd~fluxes (according to the continuum shape,
as explained in Sec.~\ref{measurement_procedures}). In
Fig.~\ref{all_plots}.d, we plot the continuum slope as defined in the
previous section vs. the \Bd/\Bg. The correlation suggests that the
origin of the continuum slope is the extinction. We transformed both
quantities in Av in Fig.~\ref{all_plots}.e. To derive the extinction
from the continuum slope, we considered that the flatter source (N2 or
source 26) has no extinction. We see that at least for the brightest
\Bg~sources (cyan dots), the values are well consistent and are in an
Av range from 10 to 30\,mag.
 
\medskip

Fig.\,\ref{all_plots} also displays three additional
plots. Plot~\ref{all_plots}.f shows that there is a possible
correlation between the width and flux of \Bg. We will not discuss any
implications of this because the apparent correlation is weak, but we
think it is worth mentioning, given the number of results in the
literature about the L-$\sigma$ relation in dwarf galaxies
\citep[][and references therein]{Bordalo11}. Plot~\ref{all_plots}.g
shows that the \HH~and \Bg~line widths are correlated, with \HH~being
narrower than \Bg~(\HH$\sim0.85$\Bg). Interestingly, in the
``filament'' joining M4 to M5, the \Bg~\fwhm~values are about 10\,\AA,
while they exceed 14\,\AA~in the region of M6. This region of higher
line width for both \Bg~and \HH~also displays a blueshift higher than
the surrounding, as can be seen in
Fig.~\ref{study_maps_big_lines}. The last plot
(Fig.~\ref{all_plots}.h) shows that the \HH~line ratio
2-1\,S(1)/1-0\,S(1) does not depend on the aperture through which it
is measured and has a value of $\sim$0.2$\pm$0.1. This is the ratio of
the fluxes without any correction for extinction. The effect of
extinction on this ratio is small since the lines are in the NIR and
their wavelengths are close. For an Av of 30\,mag, which represents
the most extreme possibility according to Fig.~\ref{all_plots}.e, an
observed ratio of 0.2 implies an actual ratio of 0.15, within the
error.

%----------------------------------------------------------------------
%----------------------------------------------------------------------

%----------------------------------------------------------------------
%                              DISCUSSION
%----------------------------------------------------------------------

\section{Discussion}
\label{discussion}

In this section, we discuss the results derived from the dataset and
presented above. The improvement of the spatial resolution with the
SINFONI data allows us to re-examine the conclusions from GA08, some
of which have been summarised in the introduction.  First, we show
that the new data allow us, contrary to GA05 and GA08, to resolve the
K-band stellar continuum emission, and hence prove that the sources
are no single clusters but rather cluster complexes. Then, we discuss
the implications of the non-detection of the CO absorption bands,
contrary to our previous work, derive new ages for the clusters, and
discuss the apparent inconsistency with our previous estimate of
7\,Myr, based on the MIR line ratios. In
Sec.~\ref{the_extended_gas_emission}, we discuss the origin of the
extended gas emission around the clusters, making full use of the
integral field capacity of SINFONI.

\subsection{Multiplicity and age}

%###########################################################################
\begin{figure}[htbp]
\begin{center}
\includegraphics[width=8cm]{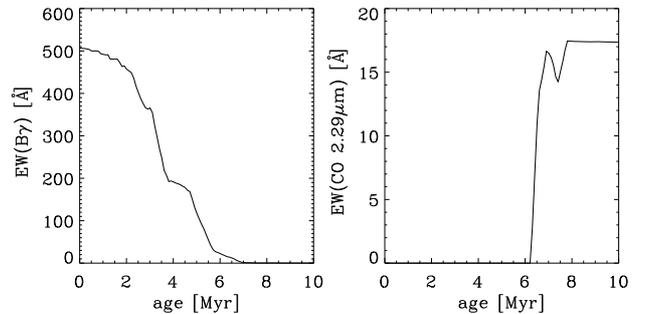}
\caption{Starburst99 equivalent widths for \Bg~emission on the left, and CO~2.29\micro~absorption band on the right.} 
\label{EWstarburst}
\end{center}
\end{figure} 
%###########################################################################

First, the SINFONI data show that M4, M5 and M6, which we repeatedly
referred to as clusters, are resolved in the K-band map at a
resolution of 0.2\arcsec, which corresponds to about 20\,pc
(Sec.~\ref{images} and Fig.~\ref{bright_dist}). Since the K-band
emission traces the stellar photospheres, these sources are very
likely cluster complexes. For M4 and M6, we do not detect multiple
sources, but resolve the peaks, hence we cannot formally refute the
possibility that they are single large ``clusters''. We note that the
K-band source M4-2, which at first sight could be considered to belong
to the M4 cluster complex, is indeed an older source, with deep CO
absorption (see next paragraph). For M5, we detected four unresolved
sources in an area of about
0.5$\times$0.5\arcsec$^2$=45$\times$45\,pc$^2$. The observational
evidence that M4, M5 and M6 are cluster complexes is the crucial new
piece of information that the SINFONI data reveal.

A second important piece of information is gained because the SINFONI
data allow us to reliably probe the presence of the CO bands at high
angular resolution, after careful evaluation of the background
emission.  Since the rapid evolution of very massive stars turns them
into red supergiants in a few million years, the NIR continuum is
dominated by them from 6-7\,Myr onwards and the deep CO bands
appear. Fig.~\ref{EWstarburst} shows the evolution predictions of
EW(CO) between 0 and 10\,Myr for a single, normal IMF, Z=0.04,
instantaneously formed cluster. It shows a steep rise between 6 and
7\,Myr. As shown by Fig. 101 and 102 of \citet{Leitherer99}, the
beginning of the steep rise of the CO depth will not depend on the
precise values of these parameters, since it only depends on the
evolution of very massive stars. For M4, M5 and M6, the measured upper
limits on EW(CO) of 2 to 4\,\AA~(Sec.~\ref{measurement_procedures} and
Fig.~\ref{all_plots}.a) formally imply an age upper limit of
6.5\,Myr. For M5-2 and M5-4, we also have a lower limit for EW(CO) and
the corresponding ages lie between 6\,Myr and 7\,Myr. According to
Starburst99, EW(CO) should never exceed
17\,\AA. Fig.~\ref{all_plots}a,b show higher values, but given the
order of magnitude of the measurement error, this is not
significant. The EW(CO) of the two brightest K-band sources (apertures
0 and 26) lie somewhere between 19 and 23\,\AA, and hence must be at
the maximum of EW(CO), corresponding to an age between 8 and 30\,Myr
for instantaneous formation, and between about 10\,Myr and
100\,Myr for continuous formation.

Another common age indicator for very young clusters is the equivalent
width of \Bg, EW(\Bg). This is valid under the hypothesis that we are
able to actually measure the continuum from the ionising cluster only,
and the \Bg~flux emitted by the gas excited by this cluster
only. Moreover, we must assume that the extinction towards the stars and
the emitting gas is the same, which is not obvious since the spatial
scale of the emitting gas is expected to be larger than that of the
stars. For this last remark, we can object that this does not seem to
be the case for the small apertures for the \Bg-emitting sources (M4,
the components of M5 and M6) since we measure a very good correlation
between the extinction derived from the K-band continuum slope and
that derived from the \Bd/\Bg~line ratio (Sec.~\ref{spectra} and
Fig.~\ref{all_plots}.e): the cyan dots lie close to the dotted
line. Fig.~\ref{all_plots}.c plots the EW(CO) vs. the EW(\Bg)
(measured following \ref{measurement_procedures}) and shows two
groups, the older sources, with low EW(\Bg) and high EW(CO), and the
younger sources with high EW(\Bg) and low EW(CO). For M4 and M6,
EW(\Bg) is about 130\,\AA, and it is about 115\,\AA~for the M5
sources. The measurement errors are small because of the measurement
procedure, but, given the uncertainties mentioned above, the real
uncertainty for the quantity that it makes sense to compare to model
predictions must be larger. Nevertheless, even if we consider a large
error of 50\,\AA, the formal ages implied by comparison with the
Starburst99 models (Fig.~\ref{EWstarburst}) lie in the narrow range
from 4.5\,Myr to 5\,Myr.

In this paragraph, we briefly state why other potentially powerful age
indicators cannot be used in our particular case. (1)
\textit{Wolf-Rayet features}: We can easily see that WR features in
the spectra of the clusters, if present, would be too faint to be
detected. To estimate the flux of a WR in the K-band, we used the WR
102ca spectrum of \citet{Homeier03}, which is is at a distance of
8.5\,kpc (quintuplet cluster) and has a flux of 0.04Jy. According to
Starburst99, the NGC1365 7Myr old clusters should contain about 1000
WR stars. At the distance of NGC1365 (18\,Mpc), this population would
have a flux of only 0.01mJy, undetectable in our data. (2)
\textit{Broad-band colours}: NGC1365 has been extensively observed by
HST, and hence images in different filters are available from the
archive, and could be used to measure the broad-band colours of the
clusters and derive ages. M4 and M6 are heavily obscured, and hence
not detected in the visible, but M5 is detected. Unfortunately, GALEV
models \citep{Kotulla09} show that we would need a 0.1mag precision on
the Av of the cluster itself to distinguish between 4Myr and 8Myr
clusters. (3) \textit{He\,I to \Bg~ratio}: In principle, this ratio
would give information about the hardness of the UV continuum emitted
by the cluster, and link it to its age \citep{Doyon92}, but this has
been demonstrated to be unreliable by \citet{Shields93} and
\citet{Lumsden03}.

In summary, for a standard instantaneously formed Starburst99 model, there is inconsistency between the different mass estimates: 

\begin{enumerate}

\item
The non-detection of the [ArIII] and [SIV] lines shows that the most
massive, hence hard UV emitting stars, have died already, hence that
the cluster is older than 7\,Myr (Fig.~15 and Fig.~16 of GA08).

\item
The non-detection of the CO absorption bands shows the absence of red
super giant stars, hence proves that the clusters are younger than
6\,Myr.

\item 
The EW(\Bg), even though we believe is not the most reliable age
indicator in our case, is consistent with ages in the range 4.5\,Myr
to 5\,Myr.

\end{enumerate}

Choosing a more complex star formation history cannot solve the above
contradiction since the age limits measured by the non-detection of
the MIR lines and CO lines are indeed the time that has elapsed after
the last massive star formation episode ended. We trust our previous
MIR observations, since we have an example of a similar or even
slightly lower quality spectrum for a comparable, but younger object,
and the [SIV] line is much brighter than [NeII]
\citep{Martin-Hernandez05}, as detailed in the introduction, and
predicted in GA08. The VISIR spectra were of good quality, several
expected spectral features are detected, and the [NeII] line has a
high S/N ratio.

An easy way of making the Starburst99 model consistent with the
observational constraints is to lower the upper mass cutoff of the IMF
($\rm M_{\max}$) to values between 25\,\msol~to 30\,\msol.  In this
case, the cluster UV continuum is simply never sufficiently hard to
maintain the ionisation of [ArIII] and [SIV].

For $\rm M_{max}$=25\,\msol~and $\rm M_{max}$=30\,Myr, we find that
the observational constraints are consistent with age ranges of
[6.2-6.4]\,Myr and [5.62-5.64]\,Myr, respectively. Hence, under the
hypothesis of a ``low'' upper mass cutoff for the IMF, the ages
derived might be only slightly lower than our first estimate in GA08,
in the range [5.5-6.5]\,Myr. The corresponding mass estimates for the
clusters would then be between 3 and 10$\times$10$^6$\,\msol. We do
not claim any evidence for a low upper mass cutoff IMF, but only show
that the observational constraints are consistent with it, and
inconsistent with the ``standard'' Starburst99 model. Given the large
masses of the these systems, the most massive stars cannot be absent
just on the basis of stochastics and an astrophysical mechanism should
be invoked to explain the possibility of an actual absence of stars
with masses greater than 25\,\msol. This is beyond the scope of the
current analysis.

In addition to the absolute ages of the cluster complexes, we note the
presence of spectral features that might witness age differences
between the clusters. We see that the slope (and extinction) increases
between M5, M4 and M6. Interestingly, in the spectrum of M6, we detected
weak [\ion{Fe}{III}] lines that may reveal a slightly harder ionising
field, hence a slightly younger age. The younger cluster complex being
the most extincted is consistent with the idea of ongoing gas
expulsion at that stage of evolution.

Even though we cannot derive any precise extinction from the SINFONI
cubes, we estimated of about 10 to 30\,mag through the small apertures
extinctions, with a good consistency between the extinction from the
gas emission lines and the star continuum (Sec.~\ref{spectra}~and
Fig.\ref{all_plots}.(e)\,). \citet{Galliano08b} derived comparatively
low extinctions towards M4, M5 and M6 of 13.5, 3.2 and 8.5\,mag,
through much broader apertures.

In summary, the new dataset has highlighted that M4, M5 and
M6 are cluster complexes, and provided evidence (EW(CO) and EW(\Bg) at
high resolution) for ages (5.5\,Myr to 6.5\,Myr) that are slightly
younger than previously derived (7\,Myr). We showed that the
inconsistency between the ages derived from the EW(CO) and EW(\Bg) on
one side, and the MIR line ratios on the other, can be overcome by
assuming an upper mass cutoff of 20 to 25\,\msol~for the IMF. We also
showed that there is some indication for an age gradient across the
sources, M6 being the youngest. The comparison between the old and the
new extinction measurements strongly suggests that the extinction is
patchy and higher towards the main sources.

\subsection{The extended gas emission} 
\label{the_extended_gas_emission}
In this subsection, we discuss the origin of the extended gas emission
around the cluster complexes.  A direct qualitative comparison between
the maps for M4, M5 and M6 and the maps available for the well-known
giant \HII~region 30 Doradus in the LMC
\citep{Indebetouw09,Pellegrini10,Pellegrini11} can help in
understanding the origin of the extended emission in the observed
field. An impressive image of the whole Tarantula nebula was published
in the ESO Photo-Release eso0216. It shows a nebula more than 300\,pc
across, which corresponds to an angular size of more than 3\arcsec~at
the distance of NGC1365. The ionisation of the 30 Doradus nebula is
dominated by photoionisation from the central cluster R136. This shows
that even though the presence of more uniformly distributed lower mass
and recently formed clusters in the region of extended gas emission
cannot be excluded, such a configuration is not necessary to
explain the extent of the emission. The region in which we observe M4,
M5 and M6 possibly also only contains three giant HII regions, each
dominated by a central cluster complex confined in a $\sim$ 50\,pc
region. That the emission peaks are located well at the
centre of the extended emission in the case of M6, and close to the
symmetry axis of the ``filament'' joining M4 to M5, strengthens the
idea that the stellar content is dominated by central cluster
complexes. The observed slight shifts in position between the emission
peaks (continuum, \Bg~and \HH) are indeed expected for a ``30
Doradus''-like complex and clumpy gas structure for the following
reason: at a resolution of about the total angular size of the
emitting region, these emissions are coincident, but the stars
(sources of continuum), the ionised gas (source of \Bg) and the warm
gas (source of \HH) occupy different regions of space, and may hence
not be spatially coincident. This can be seen in the three colour
image of R136 displayed in GA08 (Fig. 12), with the V-band emission
tracing the stars, \Ha~the ionised gas, and 8\micro~the warm
dust. When observed at a sufficiently high resolution, the difference
in position between the peaks of these components can be unveiled.

Indeed, the extended \HH~emission is well consistent with the
\HII~emission on scales of several arcsec, while they do not coincide
at a higher level of details. Fig.~\ref{apertures} clearly shows that
the \HH 1-0\,S(1) / \Bg~ratio increases with distance to the central
sources. We may wonder whether this is a sign of local excitation of
the \HH~emission. The K-band spectra show a series of \HH~lines both
for para-~and ortho-hydrogen. The excitation mechanism for these lines
can give clues about the presence of an underlying population of
smaller mass clusters. The low 2-1\,S(1)/1-0\,S(1)$\sim$0.2 ratio
observed (Fig.~\ref{all_plots}.h) usually traces thermally excited
\HH~in shocks. However, for high densities of about 10$^5$\,cm$^{-3}$
and for an intense incident FUV radiation field, collisional
de-excitation of UV-pumped molecules thermalises the lower energy
levels \citep{Burton90,Hollenbach95,Davies03,Martin-Hernandez08}.
That the central UV sources possibly maintain ionisation of hydrogen
at large distances suggests that the gas distribution is very clumpy,
hence that the density is high. The only quantitative indication of
the gas density comes from the observed ratio of the two forbidden
[\ion{Fe}{III}] lines ($^3G_5-^3H_6$ at rest 2.2183\micro~and
$^3G_4-^3H_4$ at rest 2.2427\micro) detected in the spectra of M5 and
M6 (apertures 18 and 31). We measured 0.38$\pm$0.05 for M6 and
$\le$0.5 for M5. If we compare this to the model predictions from
\citet{Keenan92} reported in \citet{Gilbert00}, we find a density of
10$^5$\,cm$^{-3}$.  This means that we cannot decide between the two
excitation mechanisms from the 2-1\,S(1)/1-0\,S(1) ratio only. As
explained in \citet{Davies03}, while the $\nu$=2 vibrational levels
are thermalised at high density, the $\nu$=3 level is expected to
retain a fluorescent population until even higher densities. In the
transitions involving the $\nu$=3 level, the data only leave upper
limits for 3-2\,S(2), 3-2\,S(4) and 3-2\,S(5) while the 3-2\,S(3) line
is blended with a skyline. For M6 (aperture 31), we found
1-0\,S(1)/3-2\,S(2) $>$ 20, and for the ``pure'' \HH~source (aperture
15), close to M5, we found a lower limit of 35. Considering a ratio
3-2\,S(2)/3-2\,S(1) equal to the standard para-ortho ratio of 1/3 (see
next paragraph), gives 1-0\,S(1)/3-2\,S(1)$>$7 and 12,
respectively. According to \citet{Martin-Hernandez08}, for densities
of about 10$^5$\,cm$^{-3}$, the fluorescent ratio retains
values lower than 8. The measured lower limits hence suggest that the
\HH~emission is thermally excited in shocks.

As discussed by \citet{Martin-Hernandez08}, an additional diagnostic
can be obtained by the measurement of the ortho-to-para ratio of the
\HH~molecules. Where shock excitation is dominant, this ratio $\Phi$,
based on the statistical weights of the nuclear spins, is 3
\citep{Smith97}. It can take other values in the case of fluorescent
excitation \citep{Sternberg99}. Following \citet{Smith97} and
\citet{Martin-Hernandez08}, we can measure the ortho-to-para ratio for
the $\nu$=1 states ($\Phi_1$) measuring the fluxes of 1-0\,S(0),
1-0\,S(1) and 1-0\,S(2). The measured values of $\Phi_1$ tend to be
similar in all apertures, and slightly lower but consistent with 3,
with a median of 2.4 and an error of $\pm$0.7. As reported by
\citet{Martin-Hernandez08}, ratios in the range 1.5-2.2 have been
measured in PDRs. Hence, the measured values are consistent with shock
excitation, but have a median that could indicate part of fluorescent
excitation.

The different diagnostics for the excitation of \HH~suggest a picture
in which shocks are the dominant mechanism, while fluorescence,
probably also present, is marginal. That shock excitation at large
distance from the central complexes is taking place might signal the
presence of a lower mass population of young clusters. However, we can
also predict the origin of these shocks in gas outflows produced by
the central complexes themselves, without the need of invoking another
``hidden'' cluster population. The blue wing of the \Bg~line detected
at the location of M4, M5 and M6 and displayed in
Fig.~\ref{line_shapes}~is possibly the evidence of such an
outflow. The comparison with line profiles extracted away from the
main sources, which are free of this asymmetry suggests that the
outflows are indeed centred on the main sources M4, M5 and M6. The
associated velocities are some hundreds of \kms, which means that
given the size of the giant \HII~regions and their age, there has been
enough time for gas heating through shocks to occur. The non-detection of
the red wing can be simply attributed to extinction.

In summary, we find that the new data are consistent with a picture in
which the central cluster complexes constitute the source of
excitation of the whole nebula through ionisation and shocks produced
by gas outflows. Even though we do not discard it, we find no need to
invoke a lower mass cluster population to interpret the data at hand.

\section{Conclusions}
\label{conclusion}

Thanks to the integral field capacity of SINFONI, along with the
improved angular resolution achieved with the adaptive optics, our
knowledge and understanding of the three bright MIR/radio 'embedded
clusters' M4, M5 and M6 located in NGC1365 nuclear region has greatly
improved.

First, the gain in angular resolution in the new K-band maps (compared
to the former ISAAC K-band maps) allowed us to resolve the
sources. This proves that they are not single clusters, but instead
cluster complexes of several tens of parsec size.

The possibility to resolve the spectral features at a high angular
resolution allowed us to perform precise measurements of both the
equivalent width of the 2.3\,\micro~CO absorption bands and that of
the \Bg~emission line, uncontaminated by background or close-by
surrounding emission. The unexpected non-detection of the CO bands
together with the measured values of EW(\Bg) suggest ages that are
younger than previously derived. These new measurements, within the
framework of a 'standard' Starburst99 cluster model, are incompatible
with the MIR line ratios discussed in our previous paper. We do not
have any reliable explanation for this apparent inconsistency yet. But
we note that it is a hint for an upper mass cutoff of the star cluster
IMF lower than usually assumed with values below 30\,\msol.  In this
case, the complexes would only be slightly younger (5.5 to 6.5\,Myr)
than derived in our former work, with corresponding masses of
[3-10]$\times$10$^6$\,\msol~per complex. If we had kept a standard
IMF, and simply discarded our previous estimate, we would have found ages of
4\,Myr, resulting in much lower masses, of about 10$^6$\,\msol.

The complete dataset available for these sources and the extended
emission around them suggests that this gas-rich region (SA07)
consists of three co-eval multiple ``30 Doradus''-type giant
\HII~regions, each powered by a central star cluster complex. There is
a weak observational indication for a slight age gradient among them,
with the youngest (M6) displaying a steeper K-band slope and the
presence of weak higher ionisation lines. A few secondary \Bg~maxima
are detected, and witness the presence of at least some lower mass
clusters or cluster complexes, but we found no compelling need to
invoke a fully populated power law cluster mass distribution to
interpret the data at hand. The angular resolution achieved is
sufficiently high to show that the distributions of the \Bg~and
\HH~line emission is not coincident, indicating a patchy gas
distribution, where the ionised gas is probably seen on the surface of
denser knots, and reveals complex kinematics (line asymmetry and broad
width). The colder \HH~emitting gas is distributed all over the
region, and appears to be predominantly thermally excited in shocks,
rather than fluorescence in photo-dissociation regions. We interpret
the observed extended blue wing in the \Bg~profiles as a signature of
gas outflows from the main cluster complexes, and suggest that these
outflows would be the natural source of shock excitation of molecular
hydrogen over the entire region.

\begin{acknowledgements}
We warmly thank the ESO staff for data acquisition in service mode and
for support in data reduction. We are also indebted to an
anomymous referee for helpful comments.
\end{acknowledgements}

%-----------------------------
%        FIRST TABLE
%-----------------------------
%\begin{landscape}
\begin{table*}[htbp]
\begin{center}
\begin{tabular}{ccccccc}

aperture &
name &
position &
K flux density&
\Bg~flux &
slope &
EW(CO) \\

\hline

\medskip

 0 &    N1 & (-0.57; 7.66) &   4.61e-01$\pm$  5.53e-03 &   1.38e-17$\pm$  7.80e-18 & 1.02$\pm$ 0.05 &  21.7$\pm$  2.2 \\
 1 & ----- & (-0.03; 7.03) &   1.89e-02$\pm$  2.10e-03 &   6.67e-16$\pm$  6.63e-18 & 1.01$\pm$ 0.10 &  17.7$\pm$  3.8 \\
 2 & ----- & (-0.03; 8.03) &   5.83e-02$\pm$  5.17e-03 &   1.40e-17$\pm$  4.81e-18 & 1.05$\pm$ 0.07 &  23.7$\pm$  3.2 \\
 5 & ----- & (0.16; 6.22) &   2.57e-02$\pm$  7.29e-04 &   1.28e-16$\pm$  3.73e-18 & 1.41$\pm$ 0.15 &   $<$ 11.7 \\
 6 &  M4-2 & (0.33; 7.33) &   6.74e-02$\pm$  1.95e-03 &   6.68e-16$\pm$  6.69e-18 & 1.12$\pm$ 0.06 &  15.9$\pm$  3.0 \\
10 &    M4 & (0.41; 7.07) &   7.98e-02$\pm$  2.02e-03 &   6.68e-16$\pm$  6.60e-18 & 1.46$\pm$ 0.07 &   $<$  4.7 \\
12 & ----- & (1.28; 7.54) &   2.04e-02$\pm$  8.74e-04 &   7.51e-18$\pm$  4.48e-18 & 1.19$\pm$ 0.16 &  20.6$\pm$  6.3 \\
14 & ----- & (2.01; 9.21) &   4.42e-02$\pm$  1.33e-03 &   5.45e-17$\pm$  5.39e-18 & 1.15$\pm$ 0.10 &  14.5$\pm$  4.4 \\
16 &  M5-4 & (2.56; 9.96) &   2.11e-01$\pm$  2.14e-03 &   1.56e-15$\pm$  1.92e-17 & 1.20$\pm$ 0.10 &   8.4$\pm$  3.4 \\
17 & ----- & (2.60;10.76) &   4.89e-02$\pm$  2.50e-03 &   3.02e-17$\pm$  1.14e-17 & 1.00$\pm$ 0.10 &  15.8$\pm$  3.8 \\
19 &    M5 & (2.81; 9.79) &   2.10e-01$\pm$  2.24e-03 &   1.57e-15$\pm$  1.84e-17 & 1.16$\pm$ 0.05 &   $<$  2.8 \\
20 &  M5-2 & (2.88;10.08) &   2.03e-01$\pm$  2.09e-03 &   1.54e-15$\pm$  1.98e-17 & 1.14$\pm$ 0.06 &   4.3$\pm$  2.8 \\
22 &  M5-3 & (3.13; 9.90) &   2.10e-01$\pm$  2.26e-03 &   1.59e-15$\pm$  1.88e-17 & 1.15$\pm$ 0.10 &   $<$  9.0 \\
24 & ----- & (3.64; 7.26) &   3.33e-02$\pm$  3.34e-03 &   2.93e-17$\pm$  5.55e-18 & 1.18$\pm$ 0.14 &  25.7$\pm$  5.7 \\
25 & ----- & (3.93; 8.06) &   2.96e-02$\pm$  2.64e-03 &   3.76e-17$\pm$  3.41e-18 & 1.07$\pm$ 0.18 &  10.4$\pm$  7.8 \\
26 & ----- & (3.94; 5.57) &   2.47e-01$\pm$  5.24e-03 &   2.01e-17$\pm$  2.39e-18 & 1.04$\pm$ 0.04 &  20.4$\pm$  2.3 \\
27 &    N2 & (3.97; 5.57) &   2.47e-01$\pm$  5.24e-03 &   1.19e-17$\pm$  2.39e-18 & 0.99$\pm$ 0.05 &  21.1$\pm$  2.3 \\
30 & ----- & (4.77; 5.66) &   7.05e-02$\pm$  3.92e-03 &   2.60e-18$\pm$  3.12e-18 & 0.92$\pm$ 0.09 &  18.5$\pm$  4.0 \\
32 &    M6 & (4.78; 6.94) &   2.60e-01$\pm$  1.40e-03 &   2.23e-15$\pm$  1.70e-17 & 1.68$\pm$ 0.04 &   $<$  1.7 \\

\end{tabular}
\caption[]{Results from the measurements described in
  Sec.~\ref{measurement_procedures} for the apertures defined in the
  K-band image. The aperture number is defined in Fig.~\ref{apertures}
  and the corresponding name is given in the second column when
  appropriate. The positions of the apertures are
  ($\Delta$(R.A.);$\Delta$(dec)) in arcsec with respect to the
  position of the AGN in the MIR (see text). The K-band flux density
  is mJy. The \Bg~flux is in \flux. These two fluxes are
  background-subtracted fluxes measured on the reconstructed images,
  through ``large'' apertures as defined at the end of
  Sec.~\ref{measurement_procedures} (aperture radius 0.2\arcsec, and
  background between 0.2\arcsec~and 0.6\arcsec). The slope is defined
  in the last paragraph of Sec.~\ref{measurement_procedures}. The
  equivalent width of the CO absorption band EW(CO) is measured
  through the small apertures following the procedure defined in
  Sec.~\ref{measurement_procedures} and is given in \AA.}
\label{table1}
\end{center}
\end{table*}
%\end{landscape}

%-----------------------------
%        SECOND TABLE
%-----------------------------

\begin{table*}[htbp]
\begin{center}
\begin{tabular}{cccccccc}

aperture &
name &
\Bd~flux&
HeI~flux&
 \Bg~flux&
$\sigma$ \Bd &
$\sigma$HeI &
$\sigma$ \Bg \\

\hline

 3 & ----- &    1.5e-15$\pm$   1.8e-16 &    3.1e-16$\pm$   4.5e-17 &    1.9e-15$\pm$   3.4e-17 &   8.1$\pm$  0.5 &   3.9$\pm$  0.3 &   4.3$\pm$  0.1 \\
 4 & ----- &    1.7e-15$\pm$   8.3e-17 &    9.0e-16$\pm$   4.3e-17 &    2.8e-15$\pm$   2.9e-17 &   5.4$\pm$  0.1 &   4.8$\pm$  0.1 &   4.0$\pm$  0.1 \\
 7 & ----- &    1.5e-15$\pm$   9.8e-17 &    8.1e-16$\pm$   4.1e-17 &    2.8e-15$\pm$   3.8e-17 &   5.2$\pm$  0.2 &   4.9$\pm$  0.1 &   4.6$\pm$  0.1 \\
 8 & ----- &    5.5e-15$\pm$   1.6e-16 &    3.6e-15$\pm$   5.0e-17 &    1.0e-14$\pm$   4.3e-17 &   5.5$\pm$  0.1 &   5.2$\pm$  0.1 &   4.9$\pm$  0.1 \\
 9 &    M4 &    6.0e-15$\pm$   1.5e-16 &    3.8e-15$\pm$   4.9e-17 &    1.0e-14$\pm$   3.8e-17 &   5.4$\pm$  0.1 &   5.1$\pm$  0.1 &   4.8$\pm$  0.1 \\
11 & ----- &    3.0e-15$\pm$   1.2e-16 &    1.5e-15$\pm$   3.7e-17 &    4.5e-15$\pm$   3.9e-17 &   4.7$\pm$  0.1 &   4.5$\pm$  0.1 &   4.2$\pm$  0.1 \\
13 & ----- &    2.0e-15$\pm$   1.1e-16 &    1.0e-15$\pm$   5.9e-17 &    3.2e-15$\pm$   5.3e-17 &   4.7$\pm$  0.1 &   4.1$\pm$  0.1 &   4.0$\pm$  0.1 \\
15 & ----- &    2.9e-15$\pm$   9.8e-17 &    1.4e-15$\pm$   4.3e-17 &    4.8e-15$\pm$   4.0e-17 &   4.2$\pm$  0.1 &   4.1$\pm$  0.1 &   4.2$\pm$  0.1 \\
16 &  M5-4 &    5.5e-15$\pm$   1.6e-16 &    2.6e-15$\pm$   5.0e-17 &    8.7e-15$\pm$   3.6e-17 &   4.8$\pm$  0.1 &   4.7$\pm$  0.1 &   4.5$\pm$  0.1 \\
18 &    M5 &    1.3e-14$\pm$   3.1e-16 &    7.6e-15$\pm$   1.1e-16 &    2.1e-14$\pm$   8.9e-17 &   4.9$\pm$  0.1 &   4.5$\pm$  0.1 &   4.3$\pm$  0.1 \\
21 &  M5-2 &    1.1e-14$\pm$   1.7e-16 &    7.2e-15$\pm$   6.3e-17 &    1.8e-14$\pm$   5.1e-17 &   4.5$\pm$  0.1 &   4.8$\pm$  0.1 &   4.8$\pm$  0.1 \\
22 &  M5-3 &    5.6e-15$\pm$   1.3e-16 &    2.6e-15$\pm$   5.9e-17 &    7.7e-15$\pm$   3.5e-17 &   4.4$\pm$  0.1 &   5.0$\pm$  0.1 &   5.0$\pm$  0.1 \\
23 & ----- &    4.8e-15$\pm$   1.3e-16 &    2.2e-15$\pm$   5.5e-17 &    6.9e-15$\pm$   2.9e-17 &   4.4$\pm$  0.1 &   4.8$\pm$  0.1 &   5.0$\pm$  0.1 \\
28 & ----- &    1.6e-15$\pm$   5.9e-17 &    7.0e-16$\pm$   3.9e-17 &    2.7e-15$\pm$   2.8e-17 &   3.2$\pm$  0.1 &   3.6$\pm$  0.1 &   4.1$\pm$  0.1 \\
29 & ----- &    1.4e-14$\pm$   1.6e-16 &    1.1e-14$\pm$   9.5e-17 &    2.8e-14$\pm$   1.8e-16 &   5.3$\pm$  0.1 &   6.7$\pm$  0.1 &   6.5$\pm$  0.1 \\
31 &    M6 &    1.4e-14$\pm$   1.6e-16 &    1.3e-14$\pm$   8.7e-17 &    3.0e-14$\pm$   1.7e-16 &   4.9$\pm$  0.1 &   6.5$\pm$  0.1 &   6.3$\pm$  0.1 \\
33 & ----- &    9.9e-15$\pm$   1.5e-16 &    6.6e-15$\pm$   9.0e-17 &    1.6e-14$\pm$   1.5e-16 &   5.8$\pm$  0.1 &   6.5$\pm$  0.1 &   6.2$\pm$  0.1 \\
34 & ----- &    2.0e-15$\pm$   4.2e-17 &    1.5e-15$\pm$   2.7e-17 &    4.5e-15$\pm$   2.5e-17 &   3.4$\pm$  0.1 &   4.8$\pm$  0.1 &   4.3$\pm$  0.1 \\
35 & ----- &    9.2e-16$\pm$   9.2e-17 &    4.8e-16$\pm$   4.2e-17 &    1.2e-15$\pm$   2.4e-17 &   6.8$\pm$  0.3 &   7.9$\pm$  0.3 &   5.3$\pm$  0.1 \\
36 & ----- &    1.1e-15$\pm$   4.0e-17 &    4.6e-16$\pm$   3.2e-17 &    1.7e-15$\pm$   1.9e-17 &   3.6$\pm$  0.1 &   4.2$\pm$  0.1 &   3.7$\pm$  0.1 \\

\end{tabular}
\caption[]{Measurements described in Sec.~\ref{measurement_procedures}, for the apertures defined on the \Bg~image and the \HH~image. The line fluxes are in \flux, and the line Gaussian $\sigma$ are in \AA.}
\label{table2}
\end{center}
\end{table*}

%-----------------------------
%        THIRD TABLE
%-----------------------------

\begin{table*}[htbp]
\begin{center}
\begin{tabular}{cccccccccc}

aperture &
name &
1-0S(3) flux&
1-0S(2) flux&
2-1S(3) flux&
1-0S(1) flux& 
$\sigma$1-0S(3) &
$\sigma$1-0S(2) &
$\sigma$2-1S(3) &
$\sigma$1-0S(1) \\

\hline

 3 & ----- &    1.2e-15$\pm$   9.3e-17 &    6.5e-16$\pm$   4.3e-17 &    1.8e-16$\pm$   2.6e-17 &    1.7e-15$\pm$   3.2e-17 &   4.5$\pm$  0.2 &   4.1$\pm$  0.1 &   3.1$\pm$  0.2 &   4.1$\pm$  0.1\\
 4 & ----- &    6.6e-16$\pm$   6.1e-17 &    4.9e-16$\pm$   3.4e-17 &    1.8e-16$\pm$   3.3e-17 &    1.2e-15$\pm$   3.2e-17 &   3.7$\pm$  0.2 &   4.1$\pm$  0.1 &   5.3$\pm$  0.5 &   3.9$\pm$  0.1\\
 7 & ----- &    6.2e-16$\pm$   7.1e-17 &    4.8e-16$\pm$   3.6e-17 &    1.5e-16$\pm$   1.9e-17 &    1.0e-15$\pm$   3.1e-17 &   3.7$\pm$  0.2 &   4.7$\pm$  0.2 &   2.3$\pm$  0.1 &   4.0$\pm$  0.1\\
 8 & ----- &    1.5e-15$\pm$   8.9e-17 &    7.5e-16$\pm$   4.4e-17 &    3.8e-16$\pm$   3.7e-17 &    2.4e-15$\pm$   3.6e-17 &   3.7$\pm$  0.1 &   3.7$\pm$  0.1 &   4.7$\pm$  0.2 &   4.1$\pm$  0.1\\
 9 &    M4 &    1.3e-15$\pm$   9.6e-17 &    7.3e-16$\pm$   4.5e-17 &    4.1e-16$\pm$   3.8e-17 &    2.2e-15$\pm$   3.5e-17 &   3.8$\pm$  0.1 &   3.7$\pm$  0.1 &   5.0$\pm$  0.2 &   4.2$\pm$  0.1\\
11 & ----- &    7.9e-16$\pm$   8.6e-17 &    5.6e-16$\pm$   3.8e-17 &    3.0e-16$\pm$   4.8e-17 &    1.3e-15$\pm$   3.1e-17 &   3.6$\pm$  0.2 &   4.6$\pm$  0.2 &   8.4$\pm$  0.7 &   4.0$\pm$  0.1\\
13 & ----- &    7.2e-16$\pm$   1.1e-16 &    5.1e-16$\pm$   1.0e-16 &    $<$   1.2e-15 &    7.2e-16$\pm$   5.4e-17 &   5.4$\pm$  0.4 &   9.2$\pm$  0.9 &   --                     &   4.9$\pm$  0.2\\
15 & ----- &    1.7e-15$\pm$   9.4e-17 &    7.7e-16$\pm$   5.2e-17 &    2.7e-16$\pm$   3.2e-17 &    2.3e-15$\pm$   3.2e-17 &   4.0$\pm$  0.1 &   4.7$\pm$  0.2 &   4.1$\pm$  0.2 &   4.1$\pm$  0.1\\
16 &  M5-4 &    1.0e-15$\pm$   1.2e-16 &    4.5e-16$\pm$   5.2e-17 &    7.6e-17$\pm$   2.7e-17 &    1.2e-15$\pm$   4.8e-17 &   3.5$\pm$  0.2 &   4.7$\pm$  0.3 &   2.3$\pm$  0.4 &   3.7$\pm$  0.1\\
18 &    M5 &    1.2e-15$\pm$   2.1e-16 &    7.4e-16$\pm$   7.2e-17 &    3.3e-16$\pm$   4.4e-17 &    1.5e-15$\pm$   5.5e-17 &   4.1$\pm$  0.4 &   5.2$\pm$  0.3 &   3.5$\pm$  0.2 &   3.4$\pm$  0.1\\
21 &  M5-2 &    9.8e-16$\pm$   1.1e-16 &    5.9e-16$\pm$   5.2e-17 &    2.8e-16$\pm$   4.0e-17 &    1.5e-15$\pm$   4.5e-17 &   3.1$\pm$  0.2 &   4.5$\pm$  0.2 &   3.6$\pm$  0.3 &   3.7$\pm$  0.1\\
22 &  M5-3 &    1.7e-15$\pm$   1.0e-16 &    7.4e-16$\pm$   4.9e-17 &    2.4e-16$\pm$   3.6e-17 &    2.1e-15$\pm$   3.6e-17 &   4.1$\pm$  0.1 &   4.2$\pm$  0.1 &   3.6$\pm$  0.3 &   4.1$\pm$  0.1\\
23 & ----- &    1.8e-15$\pm$   9.8e-17 &    7.1e-16$\pm$   4.6e-17 &    3.0e-16$\pm$   3.5e-17 &    2.2e-15$\pm$   3.2e-17 &   4.1$\pm$  0.1 &   4.1$\pm$  0.1 &   3.5$\pm$  0.2 &   4.2$\pm$  0.1\\
28 & ----- &    5.2e-16$\pm$   6.6e-17 &    3.2e-16$\pm$   3.7e-17 &    1.2e-16$\pm$   2.3e-17 &    8.5e-16$\pm$   3.2e-17 &   4.8$\pm$  0.3 &   3.7$\pm$  0.2 &   2.4$\pm$  0.2 &   4.1$\pm$  0.1\\
29 & ----- &    1.0e-15$\pm$   1.1e-16 &    7.4e-16$\pm$   6.1e-17 &    1.9e-16$\pm$   3.4e-17 &    1.8e-15$\pm$   5.1e-17 &   5.1$\pm$  0.3 &   6.9$\pm$  0.3 &   4.7$\pm$  0.4 &   5.0$\pm$  0.1\\
31 &    M6 &    8.0e-16$\pm$   1.2e-16 &    8.1e-16$\pm$   8.0e-17 &    1.8e-16$\pm$   4.8e-17 &    1.6e-15$\pm$   5.8e-17 &   4.6$\pm$  0.4 &   8.4$\pm$  0.4 &   6.0$\pm$  0.8 &   4.7$\pm$  0.1\\
33 & ----- &    1.2e-15$\pm$   1.4e-16 &    7.1e-16$\pm$   5.5e-17 &    1.5e-16$\pm$   3.4e-17 &    1.7e-15$\pm$   4.3e-17 &   6.4$\pm$  0.4 &   7.0$\pm$  0.3 &   3.4$\pm$  0.4 &   5.7$\pm$  0.1\\
34 & ----- &    5.0e-16$\pm$   4.5e-17 &    3.5e-16$\pm$   3.3e-17 &    1.2e-16$\pm$   2.7e-17 &    9.0e-16$\pm$   2.9e-17 &   3.4$\pm$  0.1 &   4.5$\pm$  0.2 &   4.0$\pm$  0.4 &   4.0$\pm$  0.1\\
35 & ----- &    9.7e-16$\pm$   6.7e-17 &    4.8e-16$\pm$   3.5e-17 &    1.7e-16$\pm$   2.9e-17 &    1.2e-15$\pm$   3.3e-17 &   5.1$\pm$  0.2 &   5.1$\pm$  0.2 &   4.9$\pm$  0.4 &   4.9$\pm$  0.1\\
36 & ----- &    6.7e-16$\pm$   4.4e-17 &    2.2e-16$\pm$   2.2e-17 &    9.9e-17$\pm$   3.0e-17 &    9.3e-16$\pm$   2.9e-17 &   4.0$\pm$  0.1 &   2.3$\pm$  0.1 &   3.3$\pm$  0.5 &   4.4$\pm$  0.1\\

\end{tabular}
\caption[]{Measurements described in Sec.~\ref{measurement_procedures}, for the apertures defined on the \Bg~image and the \HH~image. The line fluxes are in \flux, and the line Gaussian $\sigma$ are in \AA.}
\label{table3}
\end{center}
\end{table*}
%-----------------------------
%        FOURTH TABLE
%-----------------------------

\begin{table*}[htbp]
\begin{center}
\begin{tabular}{cccccccc}

aperture &
name &
 2-1S(2) flux&
 1-0S(0) flux&
 2-1S(1) flux&
$\sigma$ 2-1S(2)&
$\sigma$ 1-0S(0)&
$\sigma$ 2-1S(1)\\

\hline

 3 & ----- &    2.0e-16$\pm$   3.5e-17 &    5.7e-16$\pm$   3.7e-17 &    4.0e-16$\pm$   4.9e-17 &   5.2$\pm$  0.5 &   4.9$\pm$  0.2 &   5.2$\pm$  0.3\\
 4 & ----- &    2.0e-16$\pm$   2.9e-17 &    4.8e-16$\pm$   3.0e-17 &    3.3e-16$\pm$   5.8e-17 &   3.8$\pm$  0.3 &   3.5$\pm$  0.1 &   4.7$\pm$  0.4\\
 7 & ----- &    $<$   2.2e-16 &    3.4e-16$\pm$   2.3e-17 &    2.0e-16$\pm$   3.2e-17 &   -- &   3.3$\pm$  0.1 &   2.8$\pm$  0.2\\
 8 & ----- &    1.6e-16$\pm$   4.5e-17 &    7.3e-16$\pm$   3.6e-17 &    4.3e-16$\pm$   4.8e-17 &   5.1$\pm$  0.7 &   4.3$\pm$  0.1 &   4.3$\pm$  0.2\\
 9 &    M4 &    1.5e-16$\pm$   3.8e-17 &    7.0e-16$\pm$   4.2e-17 &    3.9e-16$\pm$   5.4e-17 &   4.5$\pm$  0.6 &   4.6$\pm$  0.1 &   4.1$\pm$  0.3\\
11 & ----- &    2.1e-16$\pm$   4.6e-17 &    3.5e-16$\pm$   3.4e-17 &    2.0e-16$\pm$   2.4e-17 &   7.1$\pm$  0.8 &   4.1$\pm$  0.2 &   2.5$\pm$  0.1\\
13 & ----- &    $<$   3.4e-16 &    $<$   1.6e-16 &    $<$   2.7e-16 &   -- &   -- &   --\\
15 & ----- &    1.9e-16$\pm$   2.6e-17 &    6.6e-16$\pm$   3.9e-17 &    4.5e-16$\pm$   3.6e-17 &   2.8$\pm$  0.2 &   4.2$\pm$  0.1 &   4.4$\pm$  0.2\\
16 &  M5-4 &    7.0e-17$\pm$   2.2e-17 &    3.9e-16$\pm$   3.6e-17 &    2.2e-16$\pm$   5.1e-17 &   2.0$\pm$  0.3 &   4.3$\pm$  0.2 &   5.0$\pm$  0.6\\
18 &    M5 &    $<$   1.9e-14 &    5.5e-16$\pm$   4.6e-17 &    4.0e-16$\pm$   6.0e-17 &   -- &   3.8$\pm$  0.2 &   3.3$\pm$  0.3\\
21 &  M5-2 &    $<$   2.2e-14 &    5.7e-16$\pm$   5.2e-17 &    2.5e-16$\pm$   4.1e-17 &   -- &   4.9$\pm$  0.2 &   2.8$\pm$  0.2\\
22 &  M5-3 &    1.3e-16$\pm$   3.2e-17 &    6.7e-16$\pm$   4.0e-17 &    3.6e-16$\pm$   3.8e-17 &   3.4$\pm$  0.4 &   4.3$\pm$  0.1 &   3.6$\pm$  0.2\\
23 & ----- &    1.6e-16$\pm$   2.9e-17 &    6.1e-16$\pm$   3.7e-17 &    3.5e-16$\pm$   3.7e-17 &   3.2$\pm$  0.3 &   4.0$\pm$  0.1 &   3.5$\pm$  0.2\\
28 & ----- &    6.7e-17$\pm$   1.9e-17 &    2.2e-16$\pm$   2.9e-17 &    1.8e-16$\pm$   3.2e-17 &   1.8$\pm$  0.3 &   3.6$\pm$  0.2 &   2.9$\pm$  0.3\\
29 & ----- &    $<$   3.6e-14 &    5.7e-16$\pm$   7.3e-17 &    3.0e-16$\pm$   5.1e-17 &   -- &   4.8$\pm$  0.3 &   3.9$\pm$  0.3\\
31 &    M6 &    $<$   3.7e-14 &    5.1e-16$\pm$   8.1e-17 &    3.2e-16$\pm$   5.1e-17 &   -- &   4.4$\pm$  0.4 &   3.6$\pm$  0.3\\
33 & ----- &    $<$   2.2e-14 &    5.2e-16$\pm$   4.2e-17 &    2.8e-16$\pm$   4.8e-17 &   -- &   5.9$\pm$  0.2 &   5.7$\pm$  0.5\\
34 & ----- &    1.2e-16$\pm$   2.9e-17 &    3.3e-16$\pm$   2.8e-17 &    3.3e-16$\pm$   5.4e-17 &   4.6$\pm$  0.6 &   3.3$\pm$  0.1 &   6.8$\pm$  0.6\\
35 & ----- &    $<$   1.1e-16 &    4.2e-16$\pm$   3.4e-17 &    1.2e-16$\pm$   2.0e-17 &   -- &   6.4$\pm$  0.3 &   1.7$\pm$  0.1\\
36 & ----- &    1.4e-16$\pm$   2.9e-17 &    3.0e-16$\pm$   2.8e-17 &    1.4e-16$\pm$   2.0e-17 &   5.2$\pm$  0.5 &   3.4$\pm$  0.2 &   2.4$\pm$  0.2\\

\end{tabular}
\caption[]{Measurements described in Sec.~\ref{measurement_procedures}, for the apertures defined on the \Bg~image and the \HH~image. The line fluxes are in \flux, and the line Gaussian $\sigma$ are in \AA.}
\label{table4}
\end{center}
\end{table*}

\bibliographystyle{aa}
%\bibliography{references}

\end{document}